\documentclass[12pt,preprint]{aastex}
\usepackage{natbib}
\usepackage{lineno}
\linenumbers

\newcommand{\ltsima} {$\; \buildrel < \over \sim \;$}
\newcommand{\gtsima} {$\; \buildrel > \over \sim \;$}
\newcommand{\lta} {\lower.5ex\hbox{\ltsima}}
\newcommand{\gta} {\lower.5ex\hbox{\gtsima}}

\newcommand{\Int} {\displaystyle \int}

\def\nicola {}

\def\apjl{Ap. J. Lett.}
\def\apjs{Ap. J. Supp.}

\def\aap{Astron. \& Astrophys.}

\slugcomment{Date: \today}

\begin{document}

\title{Prospects for GRB science with the $Fermi$ Large Area Telescope}
\author{
D.~L.~Band\altaffilmark{1,2}, 
M.~Axelsson\altaffilmark{3}, 
L.~Baldini\altaffilmark{4}, 
G.~Barbiellini\altaffilmark{5,6}, 
M.~G.~Baring\altaffilmark{7}, 
D.~Bastieri\altaffilmark{8,9}, 
M.~Battelino\altaffilmark{10}, 
R.~Bellazzini\altaffilmark{4}, 
E.~Bissaldi\altaffilmark{11}, 
G.~Bogaert\altaffilmark{12}, 
J.~Bonnell\altaffilmark{2}, 
J.~Chiang\altaffilmark{13,14}, 
J.~Cohen-Tanugi\altaffilmark{15}, 
V.~Connaughton\altaffilmark{16}, 
S.~Cutini\altaffilmark{17}, 
F. de Palma\altaffilmark{18,19},
B.~L.~Dingus\altaffilmark{20}, 
E.~do~Couto~e~Silva\altaffilmark{13}, 
G.~Fishman\altaffilmark{21}, 
A.~Galli\altaffilmark{22}, 
N.~Gehrels\altaffilmark{2,23}, 
N.~Giglietto\altaffilmark{18,19},
J.~Granot\altaffilmark{24}, 
S.~Guiriec\altaffilmark{15, 16}, 
R.~E.~Hughes\altaffilmark{25}, 
T.~Kamae\altaffilmark{13}, 
N.~Komin\altaffilmark{26,15}, 
F.~Kuehn\altaffilmark{25}, 
M.~Kuss\altaffilmark{4}, 
F.~Longo\altaffilmark{5,6,14}, 
P.~Lubrano\altaffilmark{27},
R.~M.~Kippen\altaffilmark{20}, 
M.~N.~Mazziotta\altaffilmark{19}, 
J.~E.~McEnery\altaffilmark{2}, 
S.~McGlynn\altaffilmark{10}, 
E.~Moretti\altaffilmark{5,6}, 
T.~Nakamori\altaffilmark{28}, 
J.~P.~Norris\altaffilmark{29}, 
M.~Ohno\altaffilmark{30},
M.~Olivo\altaffilmark{5}, 
N.~Omodei\altaffilmark{4,14}, 
V.~Pelassa\altaffilmark{15}, 
F.~Piron\altaffilmark{15}, 
R.~Preece\altaffilmark{16}, 
M.~Razzano\altaffilmark{4}, 
J.~J.~Russell\altaffilmark{13}, 
F.~Ryde\altaffilmark{10}, 
P.~M.~Saz~Parkinson\altaffilmark{31}, 
J.~D.~Scargle\altaffilmark{32}, 
C.~Sgr\`o\altaffilmark{4}, 
T.~Shimokawabe\altaffilmark{28}, 
P.~D.~Smith\altaffilmark{25}, 
G.~Spandre\altaffilmark{4}, 
P.~Spinelli\altaffilmark{18,19},
M.~Stamatikos\altaffilmark{2}, 
B.~L.~Winer\altaffilmark{25}, 
R.~Yamazaki\altaffilmark{33}
}

\altaffiltext{1}{Center for Research and Exploration in Space Science and Technology (CRESST), NASA Goddard Space Flight Center, Greenbelt, MD 20771}
\altaffiltext{2}{NASA Goddard Space Flight Center, Greenbelt, MD 20771}
\altaffiltext{3}{Stockholm Observatory, Albanova, SE-106 91 Stockholm, Sweden}
\altaffiltext{4}{Istituto Nazionale di Fisica Nucleare, Sezione di Pisa, I-56127 Pisa, Italy}
\altaffiltext{5}{Istituto Nazionale di Fisica Nucleare, Sezione di Trieste, I-34127 Trieste, Italy}
\altaffiltext{6}{Dipartimento di Fisica, Universit\`a di Trieste, I-34127 Trieste, Italy}
\altaffiltext{7}{Rice University, Department of Physics and Astronomy, MS-108, P. O. Box 1892, Houston, TX 77251, USA}
\altaffiltext{8}{Istituto Nazionale di Fisica Nucleare, Sezione di Padova, I-35131 Padova, Italy}
\altaffiltext{9}{Dipartimento di Fisica ``G. Galilei", Universit\`a di Padova, I-35131 Padova, Italy}
\altaffiltext{10}{Department of Physics, Royal Institute of Technology (KTH), AlbaNova, SE-106 91 Stockholm, Sweden}
\altaffiltext{11}{Max-Planck Institut f\"ur extraterrestrische Physik, Giessenbachstra\ss e, 85748 Garching, Germany}
\altaffiltext{12}{Laboratoire Leprince-Ringuet, \'Ecole polytechnique, CNRS/IN2P3, Palaiseau, France}
\altaffiltext{13}{W. W. Hansen Experimental Physics Laboratory, Kavli Institute for Particle Astrophysics and Cosmology, Department of Physics and Stanford Linear Accelerator Center, Stanford University, Stanford, CA 94305}
\altaffiltext{14}{Corresponding authors: J.~Chiang, jchiang@slac.stanford.edu; F.~Longo, francesco.longo@trieste.infn.it; N.~Omodei, nicola.omodei@pi.infn.it.}
\altaffiltext{15}{Laboratoire de Physique Th\'eorique et Astroparticules, Universit\'e Montpellier 2, CNRS/IN2P3, Montpellier, France}
\altaffiltext{16}{University of Alabama in Huntsville, Huntsville, AL 35899}
\altaffiltext{17}{Agenzia Spaziale Italiana (ASI) Science Data Center, I-00044 Frascati (Roma), Italy}
\altaffiltext{18}{Dipartimento di Fisica ``M. Merlin" dell'Universit\`a e del Politecnico di Bari, I-70126 Bari, Italy}
\altaffiltext{19}{Istituto Nazionale di Fisica Nucleare, Sezione di Bari, 70126 Bari, Italy}
\altaffiltext{20}{Los Alamos National Laboratory, Los Alamos, NM 87545, USA}
\altaffiltext{21}{NASA Marshall Space Flight Center, Huntsville, AL 35805}
\altaffiltext{22}{INAF-Istituto di Astrofisica Spaziale e Fisica Cosmica, I-00133 Roma, Italy}
\altaffiltext{23}{University of Maryland, College Park, MD 20742}
\altaffiltext{24}{Centre for Astrophysics Research, University of Hertfordshire, College Lane, Hatfield AL10 9AB}
\altaffiltext{25}{Department of Physics, Center for Cosmology and Astro-Particle Physics, The Ohio State University, Columbus, OH 43210}
\altaffiltext{26}{Laboratoire AIM, CEA-IRFU/CNRS/Universit\'e Paris Diderot, Service d'Astrophysique, CEA Saclay, 91191 Gif sur Yvette, France}
\altaffiltext{27}{Istituto Nazionale di Fisica Nucleare, Sezione di Perugia, I-06123 Perugia, Italy}
\altaffiltext{28}{Department of Physics, Tokyo Institute of Technology, Meguro City, Tokyo 152-8551, Japan}
\altaffiltext{29}{Department of Physics and Astronomy, University of Denver, Denver, CO 80208}
\altaffiltext{30}{Institute of Space and Astronautical Science, JAXA, 3-1-1 Yoshinodai, Sagamihara, Kanagawa 229-8510, Japan}
\altaffiltext{31}{Santa Cruz Institute for Particle Physics, Department of Physics and Department of Astronomy and Astrophysics, University of California at Santa Cruz, Santa Cruz, CA 95064}
\altaffiltext{32}{Space Sciences Division, NASA Ames Research Center, Moffett Field, CA 94035-1000}
\altaffiltext{33}{Department of Physical Science and Hiroshima Astrophysical Science Center, Hiroshima University, Higashi-Hiroshima 739-8526, Japan}

\begin{abstract}

The LAT instrument on the $Fermi$ mission will reveal the rich
spectral and temporal gamma-ray burst phenomena in the $>$100~MeV
band. The synergy with $Fermi$'s GBM detectors will link these
observations to those in the well-explored 10--1000~keV range; the
addition of the $>$100~MeV band observations will resolve
theoretical uncertainties about burst emission in both the prompt
and afterglow phases. Trigger algorithms will be applied to the LAT
data both onboard the spacecraft and on the ground. The sensitivity
of these triggers will differ because of the available computing
resources onboard and on the ground.  Here we present the LAT's
burst detection methodologies and the instrument's GRB capabilities.

\end{abstract}


%
%

\section{Introduction}\label{sec:Introduction}

\begin{figure}
  \plotone{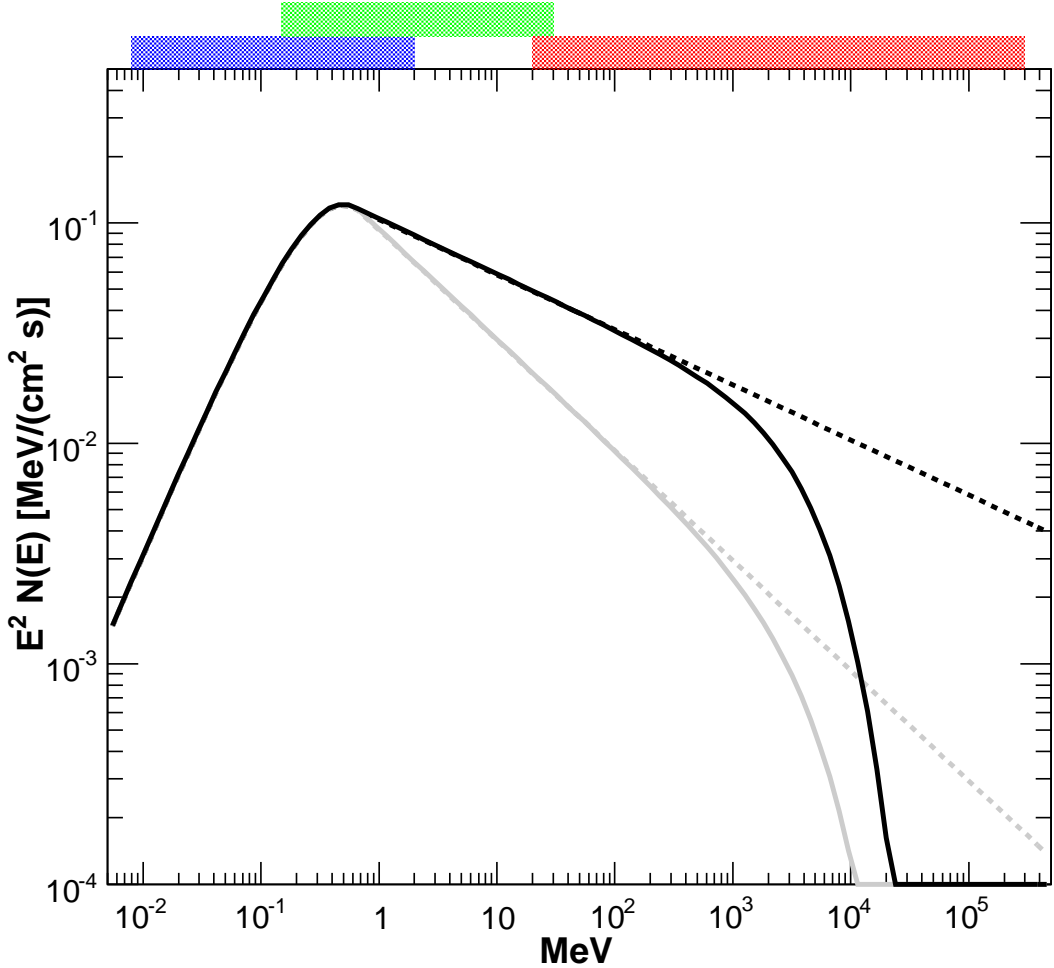}
  \caption{Simulated gamma-ray burst spectra, showing the broad
  energy range covered by $Fermi$: (from left to right) the GBM
  NaI (blue band:  8--2000~keV), the GBM BGO (green:
  150~keV--30~MeV) and the LAT (red curve: 20~MeV to $>$300~GeV)
  detectors.  The
  dashed curves are simple extrapolations of the typical GRB
  10--1000~keV spectra into the GeV~band, while the solid
  curves add an
  exponential cutoff that might result from absorption internal
  or external to the burst.  The two different high energy
  photon indices $\beta$=-2.25 (black curves) and $\beta$=-2.5
  (grey curves) demonstrate the dependence of the expected LAT
  flux on this photon index.  There may be additional high energy
  components that are not known yet and are not shown in the figure.}
  \label{fig:SP_1}
\end{figure}

The Large Area Telescope (LAT) on the $Fermi$ Gamma-ray Space
Telescope (formerly GLAST---Gamma-ray Large Area Space Telescope)
will turn the study of the 20~MeV to more than 300~GeV spectral and
temporal behavior of gamma-ray bursts (GRBs) from speculation based
on a few suggestive observations to a decisive diagnostic of the
emission processes. The burst observations of the Energetic
Gamma-Ray Experiment Telescope (EGRET) on the {\it Compton Gamma-Ray
Observatory (CGRO)} suggested three types of high energy emission:
an extrapolation of the 10--1000~keV spectral component to the
$>$100~MeV band; an additional spectral component during the
$<$1~MeV `prompt' emission; and high energy emission that lingers
long after the prompt emission has faded away. The LAT's
observations, in conjunction with the Gamma-ray Burst Monitor
(GBM---8~keV to 30~MeV), will provide unprecedented
spectral-temporal coverage for a large number of bursts. The spectra
from these two instruments will cover seven and a half energy
decades ($<$10~keV to $>$300~GeV; see Fig.~\ref{fig:SP_1}, which
shows different theoretically-predicted spectra). Thus the LAT will
explore the rich phenomena suggested by the EGRET observations,
probing the physical processes in the extreme radiating regions.

In this paper we provide the scientific community interested in GRBs
with an overview of the LAT's operations and capabilities in this
research area. Our development of detection and analysis tools has
been guided by the previous observations and the theoretical
expectations for emission in the $>$100~MeV band
(\S~\ref{sec:Expectations}).  The LAT is described in depth in an
instrument paper\citep{Atwood:09}, and therefore
here we only provide a brief summary of the $Fermi$ mission and the
LAT, focusing on issues relevant to burst detection and analysis
(\S~\ref{sec:Fermi_Mission}). Simulations are the basis of our
analysis of the mission's burst sensitivity, and are largely based
on {\it CGRO} observations (\S~\ref{sec:Burst_Simulations}). We use
our simulation methodology to estimate the ultimate burst
sensitivity and the resulting burst flux distribution
(\S~\ref{sec:Design_Sensitivity}). Both the LAT and the GBM will
apply burst detection algorithms onboard and on the ground, and the
efficiency of these methods will determine which bursts the LAT will
detect, and with what latency (\S~\ref{sec:GRB_Detection}). Once a
burst has been detected, spectral and temporal analysis of LAT (and
GBM) data will be possible (\S~\ref{sec:Spectral_Analysis}). The
burst observations by ground-based telescopes and other space
missions, particularly {\it Swift}, will complement the $Fermi$
observations (\S~\ref{sec:Other_Missions}). While basic methods are
in place for detecting and analyzing burst data, in-flight
experience will guide future work (\S~\ref{sec:Conclusions}).

%
%
%

%
%

\section{Burst Physics Above 100 MeV}\label{sec:Expectations}

\subsection{Previous Observations}\label{sec:Previous_Observations}

The detectors of the {\it Compton Gamma-Ray Observatory (CGRO)}
provided time-resolved spectra for a statistically well-defined
burst population.  These observations are the foundation of our
expectations for $Fermi$'s discoveries, which have guided the
development of analysis tools before launch.

The Burst And Transient Source Experiment (BATSE) on {\it CGRO}
observed a large sample of bursts in the $\sim$25--2000~keV band
with well-understood population statistics \citep{Paciesas:99}.
Spectroscopy by the BATSE detectors found that the emission in this
energy band could be described by the empirical four parameter
``Band'' function \citep{Band:93}
\begin{equation}
\label{eqn:BandF2}
N_{\rm Band}(E | N_0,E_p,\alpha,\beta) = N_0 \left\{
\begin{array}{l l}
   E^{\alpha} \exp[-E(2+\alpha)/E_p], &  E \le {{\alpha-\beta}\over {2+\alpha}} E_p \\
\\
   E^\beta \left[ {{\alpha-\beta}\over {2+\alpha}} E_p\right]^{(\alpha-\beta)}
   \exp[\beta-\alpha] , & E > {{\alpha-\beta}\over {2+\alpha}} E_p,
\end{array}
\right.
\end{equation}
where $\alpha$ and $\beta$ are the low and high energy
photon indices, respectively, and $E_p$ is the `peak
energy' which corresponds to the maximum of $E^2 N(E)
\propto \nu f_\nu$ for the low energy component. Typically
$\alpha \sim -0.5$ to $-1$ and $\beta$ is less than $-2$
\citep{Band:93,Preece:00,Kaneko:06}; the total energy would
be infinite if $\beta \geq -2$ unless the spectrum has a
high energy cutoff. The observations of 37 bursts by the
Compton Telescope (COMPTEL) on {\it CGRO} (0.75--30~MeV)
are consistent with the BATSE observations of this spectral
component \citep{Hoover:05}. Because of the relatively poor
spectral resolution of the BATSE detectors
\citep{Briggs:99}, this functional form usually is a good
description of spectra accumulated over both short time
periods and entire bursts, even though bursts show strong
spectral evolution \citep{Ford:95}.  It is this
10--1000~keV `prompt' component that is well-characterized
and therefore provides a basis for quantitative
predictions. A detailed duration-integrated spectral
analysis (in ~30 keV-200 MeV) of the prompt emission for 15
bright BATSE GRB performed by \citet{Kaneko:08} confirmed
that only in few case there's a significant high-energy
excess with respect to low energy spectral extrapolations.


The burst observations by the Energetic Gamma-Ray
Experiment Telescope (EGRET) on {\it CGRO} (20~MeV to
30~GeV) provide the best prediction of the LAT
observations.  EGRET observed different types of high
energy burst phenomena. Four bursts had simultaneous
emission in both the EGRET and BATSE energy bands,
suggesting that the spectrum observed by BATSE extrapolates
to the EGRET energy band \citep{Dingus:03}.  However, the
correlation with the prompt phase pulses was hampered by
the severe EGRET spark chamber dead time
($\sim$100~ms/event) that was comparable or longer than the
pulse timescales.  The EGRET observations of these bursts
suggest that the $\sim$1~GeV emission often lasts longer
than the lower energy emission, and thus results in part
from a different physical origin. 
A similar behaviour is present also in GRB 080514B detected by AGILE\citep{Giuliani:08}.

Whether high energy emission is present in both long and
short bursts is unknown. The four bursts with high energy
emission detected by EGRET were all long bursts, although
GRB 930131 is an interesting case. It was detected by BATSE
\citep{Kouveliotou:94} with duration of $T_{90}$=14~s
\footnote{$T_{90}$ is the time over which 90\% of the
emission occurs in a specific energy band.} and found to
have high-energy ($>$30~MeV) photons accompanying the
prompt phase and possibly extending beyond
\citep{Sommer:94}.  The BATSE lightcurve is dominated by a
hard initial emission lasting ~1 sec and followed by a smooth
extended emission. This burst may, therefore, have been one
of those long bursts possibly associated with a merger and
not a collapsar origin, commonly understood as the most
probable origin for short and long burst
respectively\citep{Zhang:07}. Several events have now been
identified that could fit into this category
\citep{Norris:06} and their origin is still uncertain. LAT
will make an important contribution in determining the
nature of the high energy emission from similar events and
a larger sample of bursts with detected high energy
emission will determine whether the absence of high energy
emission differentiates short from long bursts.


A high energy temporally resolved spectral component in
addition to the Band function is clearly present in GRB
941017 \citep{Gonzalez:03}; this component is harder than
the low energy prompt component, and continues after the
low energy component fades into the background.  The time
integrated spectra of both GRB~941017 and GRB~980923 show
this additional spectral component \citep{Kaneko:08}.

Finally, the $>$1~GeV emission lingered for 90~minutes
after the prompt low energy emission for GRB~940217,
including an 18~GeV photon 1.5 hours after the burst
trigger \citep{Hurley:94}.  Whether this emission is
physically associated with the lower energy afterglows is
unknown.

These three empirical types of high energy emission---an
extrapolation of the low energy spectra; an additional spectral
component during the low energy prompt emission; and an
afterglow---guide us in evaluating $Fermi$'s burst observation
capabilities.

Because the prompt low energy component was characterized
quantitatively by the BATSE observations while the EGRET
observations merely demonstrated that different components were
present, our simulations are based primarily on extrapolations of
the prompt low energy component from the BATSE band to the
$>$100~MeV band.  We recognize that the LAT will probably detect
additional spectral and temporal components, or spectral cutoffs,
that are not treated in this extrapolation.

During the first few months of the $Fermi$ mission, LAT detected already emission from three GRBs: 080825C \citep{Bouvier:08}, 080916C \citep{Tajima:08} and 081024B \citep{Omodei:08}. The rich phenomenology of high energy emission is confirmed in these three events, where spectral measurements over various order of magnitude were possible together with the detection of extended emission and spectral lags. 
In particular, the GRB 080916C was bright enough to afford unprecedented broad-band spectral coverage in four distinct time intervals \citep{Abdo:09}, thereby offering new insights into the character of energetic bursts.

\subsection{Theoretical Expectations}\label{sec:Theory}

In the current standard scenario, the burst emission arises
in a highly relativistic, unsteady outflow.  Several
different progenitor types could create this outflow, but
the initial high optical depth within the outflow obscures
the progenitor type. As this outflow gradually becomes
optically thin, dissipation processes within the outflow,
as well as interactions with the surrounding medium, cause
particles to be accelerated to high energies and loose some
of their energy into radiation. Magnetic fields at the
emission site can be strong and may be caused by a
frozen-in component carried out by the outflow from the
progenitor, or may be built up by turbulence or
collisionless shocks. The emitted spectral distribution
then depends on the details of the radiation mechanism,
particle acceleration, and the dynamics of the explosion
itself.

`Internal shocks' result when a faster region catches up with a
slower region within the outflow. `External shocks' occur
at the interface between the outflow and the ambient
medium, and include a long-lived forward shock that is
driven into the external medium and a short-lived reverse
shock that decelerates the outflow.  Thus the simple model
of a one-dimensional relativistic outflow leads to a
multiplicity of shock fronts, and many possible interacting
emission regions.

As a result of the limited energy ranges of past and current
experiments, most theories have not been clearly and unambiguously
tested. $Fermi$'s GBM and LAT will provide an energy range broad
enough to distinguish between different origins of the emission; in
particular the unprecedented high-energy spectral coverage will
constrain the total energy budget and radiative efficiency, as
potentially most of the energy may be radiated in the LAT range. The
relations between the high and low energy spectral components can
probe both the emission mechanism and the physical conditions in the
emission region. The shape of the high energy spectral energy
distribution will be crucial to discriminate between hadronic
cascades and leptonic emission.  The spectral breaks at high energy
will constrain the Lorentz factor of the emitting region. Previously
undetected emission components might be present in the light curves
such as thermal emission. Finally, temporal analysis of the high
energy delayed component will clarify the nature of the flares seen
in the X-ray afterglows.

\subsubsection{Leptonic vs. Hadronic Emission
Models}\label{sec:Leptonic_Hadronic}

It is very probable that particles are accelerated to very
high energies close to the emission site in GRBs. This
could either be in shock fronts, where the Fermi mechanism
or other plasma instabilities can act, or in magnetic
reconnection sites. Two major classes of
models---synchrotron and inverse Compton emission by
relativistic electrons and protons, and hadronic
cascades---have been proposed for the conversion of
particle energy into observed photon radiation.

In the leptonic models, synchrotron emission by relativistic
electrons can explain the 10~keV--1~MeV spectrum in $\sim$2/3 of
bursts (e.g., see \citealt{Preece:98}), and inverse Compton (IC)
scattering of low energy seed photons generally results in GeV band
emission.  These processes could operate in both internal and
external shock regions (see, e.g., \citealt{Zhang:01}), with the
relativistic electrons in one region scattering the `soft' photons
from another region
\citep{Fragile:04,Fan:05,Meszaros:94,Waxman:97,Panaitescu:98}.
Correlated high and low energy emission is expected if the same
electrons radiate synchrotron photons and IC scatter soft photons.
In Synchrotron Self-Compton (SSC) models the electrons' synchrotron
photons are the soft photons and thus the high and low energy
components should have correlated variability
\citep{Guetta:03,Galli:08}. However, SSC models tend to generate
a broad $\nu F_{\nu}$ peak in the MeV band, and for bursts observed
by {\it CGRO} this breadth has difficulty accommodating the observed
spectra \citep{Baring:04}. $Fermi$, with its broad spectral coverage
enabled by the GBM and the LAT, is ideally suited for probing this
issue further.

Alternatively, photospheric thermal emission might dominate
the soft keV--MeV range during the early part of the prompt
phase \citep{Rees:05,Ryde:04,Ryde:05}.  Such a component is
expected when the outflow becomes optically thin, and would
explain low energy spectra that are too hard for
conventional synchrotron models
\citep{Crider:97,Preece:98,Preece:02}. An additional power
law component might underlie this thermal component and
extend to high energy; this component might be synchrotron
emission or IC scattering of the thermal photons by
relativistic electrons.  Fits of the sum of thermal and
power law models to BATSE spectra have been successful
\citep{Ryde:04,Ryde:05}, but joint fits of spectra from the
two types of GBM detectors and the LAT should resolve
whether a thermal component is present
\citep{Battelino:07a, Battelino:07b}.

In hadronic models relativistic protons scatter inelastically off
the $\sim$100~keV burst photons ($p\gamma$ interactions) producing
(among other possible products) high-energy, neutral pions ($\pi^0$)
that decay, resulting in gamma rays and electrons that then radiate
additional gamma rays. Similarly, if neutrons in the outflow
decouple from protons, inelastic collisions between neutrons and
protons can produce pions and subsequent high energy emission
\citep{Derishev:00,Bahcall:00}. High energy neutrinos that may be
observable are also emitted in these interactions
\citep{Waxman:97a}.  Many variants of hadronic cascade models have
been proposed: high energy emission from proton-neutron inelastic
collisions early in the evolution of the fireball
\citep{Bahcall:00}; proton-synchrotron and photo-meson cascade
emission in internal shocks (e.g.,
\citealt{Totani:98,Zhang:01,Fragile:04,Gupta:07}); and proton
synchrotron emission in external shocks \citep{Bottcher:98}.  A
hadronic model has been invoked to explain the additional spectral
component observed in GRB~941017 \citep{DermerAtoyan:04}. The
emission in these models is predicted to peak in the MeV to GeV band
\citep{Bottcher:98,Gupta:07}, and thus would produce a clear signal
in the LAT's energy band. However, photon-meson interactions would
result from a radiatively inefficient fireball \citep{Gupta:07},
which is in contrast with the high radiative efficiency that is
suggested by {\it Swift} observations \citep{Nousek:06,Granot:06}.
Thus, the hadronic mechanisms for gamma-ray production
are many, but the $Fermi$ measurements of the temporal evolution of
the highest energy photons will provide strong constraints on these
models, and moreover discern the existence or otherwise of distinct GeV-band components.

\subsubsection{High-Energy Absorption}\label{sec:HE_abs}

At high energies the outflow itself can become optically
thick to photon-photon pair production, causing a break in
the spectrum. Signatures of internal absorption will
constrain the bulk Lorentz factor and adiabatic/radiative
behavior of the GRB blast wave as a function of time
\citep{Baring:97,Lithwick:01,Guetta:03,Baring:06,Granot:08}.
Since the outflow might not be steady and may evolve during
a burst, the breaks should be time-variable, a distinctive
property of internal attenuation.  Moreover, if the
attenuated photons and their hard X-ray/soft gamma-ray
target photons originate from proximate regions in the
bursts, the turnovers will approximate broken power-laws. Interestingly, 
the LAT has already provided palpable new
advances in terms of constraining bulk motion in bursts.  For
GRB 080916C, the absence of observable attenuation turnovers up
to around 13 GeV suggests that the bulk Lorentz factor may be
well in excess of 500-800 \citep{Abdo:09}.

Spectral cutoffs produced by internal absorption must be
distinguished observationally from cutoffs caused by interactions
with the extragalactic background. The optical depth of the Universe
to high-energy gamma rays resulting from pair production on infrared
and optical diffuse extragalactic background radiation can be
considerable, thereby preventing the radiation from reaching us.
These intervening background fields necessarily generate
quasi-exponential turnovers familiar to TeV blazar studies, which
may well be discernible from those resulting from internal
absorption. Furthermore, their turnover energies should not vary
with time throughout the burst, another distinction between the two
origins for pair attenuation. In addition, the turnover energy for
external absorption is expected above a few 10's of GeV while for
internal absorption it may be as low as $\lesssim 1\;$GeV
\citep{Granot:08}. Although the external absorption may complicate
the study of internal absorption, studies of the cutoff as a
function of redshift can measure the universe's optical energy
emission out to the Population~III epoch (with redshift $> 7$)
\citep{deJager:02,Coppi:97,Kashlinsky:05,BrommLoeb:06}.

\subsubsection{Delayed GeV Emission}

The observations of GRB~940217 \citep{Hurley:94}
demonstrated the existence of GeV-band emission long after
the $\sim$100~keV `prompt' phase in at least some bursts.
With the multiplicity of shock fronts and with synchrotron
and IC components emitted at each front, many models for
this lingering high energy emission are possible. In
combination with the prompt emission observations and
afterglow observations by {\it Swift} and ground-based
telescopes, the LAT observations may detect spectral and
temporal signatures to distinguish between the different
models.

These models include:  Synchrotron Self-Compton (SSC)
emission in late internal shocks (LIS)
\citep{Zhang:02,Wang:06,Fan:08,Galli:08}; external IC (EIC)
scattering of LIS photons by the forward shock electrons
that radiate the afterglow \citep{Wang:06}; IC emission in
the external reverse shock (RS)
\citep{Wang:01,Granot:03,Kobayashi:07}; and SSC emission in
forward external shocks
\citep{MeszarosRees:94,Dermer:00b,Zhang:01,Dermer:07,Galli:07}.

A high energy IC component may be delayed and have broader
time structures relative to lower energy components because
the scattering may occur in a different region from where
the soft photons are emitted \citep{Wang:06}.  The
correlation of GeV emission with X-ray afterglow flares
observed by {\it Swift} would be a diagnostic for different
models \citep{Wang:06,Galli:07,Galli:08}.

\subsection{Timing Analysis}

The LAT's low deadtime and large effective area will permit
a detailed study of the high energy GRB light curve, which
was impossible with the EGRET data as a result of the large
deadtime that was comparable to typical widths of the peaks
in the lightcurve. These measures are clearly important for
determining the emission region size and the Lorentz factor
in the emitting fireball.

The lightcurves of GRBs are frequently complex and diverse.
Individual pulses display a hard-to-soft evolution, with
$E_p$ decreasing exponentially with the burst flux. One
method of classifying bursts is to examine the spectral
lag, which relates to the delay in the arrival of high
energy and low energy photons
\citep[e.g.,][]{Norris:00,Foley:08}. A positive lag value
indicates hard-to-soft evolution
\citep{Kocevski:03,Hafizi:07}, i.e., high energy emission
arrives earlier than low energy emission. This lag is a
direct consequence of the spectral evolution of the burst
as $E_p$ decays with time. The distributions of spectral
lags of short and long GRBs are noticeably different, with
the lags of short GRBs concentrated in the range $\pm$
30\,ms \citep[e.g.,][]{Norris:06,Yi:06}, while long GRBs
have lags covering a wide range with a typical value of
100\,ms \citep[e.g.,][]{Hakkila:07}. \citet{Stamatikos:08}
study the spectral lags in the {\it Swift} data.

An anti-correlation has been discovered between the lag and the peak
luminosity of the GRB at energies $\sim$ 100\,keV \citep{Norris:00},
using six BATSE bursts with definitive redshift. Brighter long GRBs
tend to have a high peak luminosity and short lag, while weaker GRBs
tend to have lower luminosities and longer lags. This
``lag--luminosity relation'' has been confirmed by using a number of
\textit{Swift} GRBs with known redshift \citep[e.g., GRB~060218,
with a lag greater than 100~s, ][]{Liang:06}. $Fermi$ will be able to
determine if this relation extends to MeV-GeV energies.

A subpopulation of local, faint, long lag GRBs has been proposed by
\citet{Norris:02} from a study of BATSE bursts, which implies that
events with low peak fluxes ($F_{P} (50-300\hbox{ keV}) \sim
0.25$\,ph\,cm$^{-2}$\,s$^{-1}$) should be predominantly long lag
GRBs. \citet{Norris:02} successfully tested a prediction that these
long lag events are relatively nearby and show some spatial
anisotropy, and found a concentration towards the local
supergalactic plane. This has been confirmed with the GRBs observed
by INTEGRAL \citep{Foley:08}  where it was found that $>$\,90\% of
the weak GRBs with a lag $>0.75$\,s were concentrated in the
supergalactic plane\footnote{A possible
 counterargument has been recently claimed by \cite{xiao-schaefr}}. $Fermi$ measures of long lag GRBs will confirm
this hypothesis.  An underluminous abundant population is inferred
from observations of nearby bursts associated with supernovae
\citep{Soderberg:06}.

Moreover, some Quantum Gravity (QG) theories predict an
energy dependent speed-of-light (see e.g.,
\citealt{Mattingly:05}), which is often parameterized as
\begin{equation}
v = c\left(1-\left(E(z)/E_{qg}\right)\right)
\end{equation}
where $E(z)$ is the photon energy at a given redshift, $E(z)=
E_{obs}(1+z)$, and $E_{qg}$ is the QG scale, which may be of order
$\sim 10^{19}$~GeV.  This energy-dependence can be measured from the
difference in the arrival times of different-energy photons that
were emitted at the same time; measurements thus far give $E_{qg}$
greater than a few times $10^{17}$~GeV.  Such photons might be
emitted in sharp burst pulses \citep{Amelino-Camelia:98};
measurements have been attempted \citep{Schaefer:99,Boggs:04}. 
The most difficult roadblock to reliable quantum gravity
detections or upper limits results from the difficulty in
discriminating against time delays inherent in the emission
at the site of the GRB itself, and known to exist from previous
observations.  This problem can be addressed by studying
a sample of bursts at different redshifts, or otherwise
calibrating this effect.

With the energy difference between the GBM's low energy end and the LAT's
high energy end, the good event timing by both the GBM and the LAT,
and the LAT's sensitivity to high energy photons, the $Fermi$ mission
will place interesting limits on $E_{qg}$.

%

\section{Description of the $Fermi$ Mission}\label{sec:Fermi_Mission}

\subsection{Mission Overview}

$Fermi$ was launched on June 11, 2008, into a 96.5~min circular orbit
565~km above the Earth with an inclination of 25.6$^\circ$ to the
Earth's equator. During the South Atlantic Anomaly passages
(approximately 17\% of the time, on average) the $Fermi$ detectors do
not take scientific data. In $Fermi$'s default observing mode the
LAT's pointing is offset 35$^\circ$ from the zenith direction
perpendicular to the orbital plane; the pointing will be rocked from
one side of the orbital plane to the other once per orbit. This
observing pattern results in fairly uniform LAT sky exposure over
two orbits; the uniformity is increased by the 54~d precession of
the orbital plane.

The mission's telemetry is downlinked 6--8 times per day on
the Ku band through the Tracking and Data Relay Satellite
System (TDRSS).\footnote{See
http://msl.jpl.nasa.gov/Programs/tdrss.html}  The time
between these downlinks, the transmission time through
TDRSS and the processing at the LAT Instrument Science and
Operations Center (LISOC) result in a latency of 6 hours
between an observation and the availability of the
resulting LAT data for astrophysical analysis.  In
addition, when burst detection software for either detector
triggers, messages are sent to the ground through TDRSS
with a $\sim$15~s latency. The mission's burst operations
are described in greater detail below.

\subsection{The Large Area Telescope (LAT)}\label{sec:LAT_Description}

A product of an international collaboration between NASA,
DOE and many scientific institutions across France, Italy,
Japan and Sweden, the LAT is a pair conversion telescope
designed to cover the energy band from 20~MeV to greater
than 300~GeV.  The LAT is described in greater depth in \cite{Atwood:09} 
and here we summarize
salient features useful for understanding the detector's
burst capabilities.  The LAT consists of an array of $4
\times 4$ modules, each including a tracker-converter based
on Silicon Strip Detector (SSD) technology and a 8.5
radiation lengths CsI hodoscopic calorimeter. High energy
incoming gamma-rays convert into electron-positron pairs in
one of the tungsten layers that are interleaved with the
SSD planes; the pairs are then tracked to point back to the
original photons' direction and their energy is measured by
the calorimeter. A segmented anti-coincident shield
surrounding the whole detector ensures the necessary
background rejection power against charged particles, whose
flux outnumbers that of gamma-rays by several orders of
magnitude, and reduce the data volume to fit in the
telemetry bandwidth.

\begin{figure}
  \plottwo{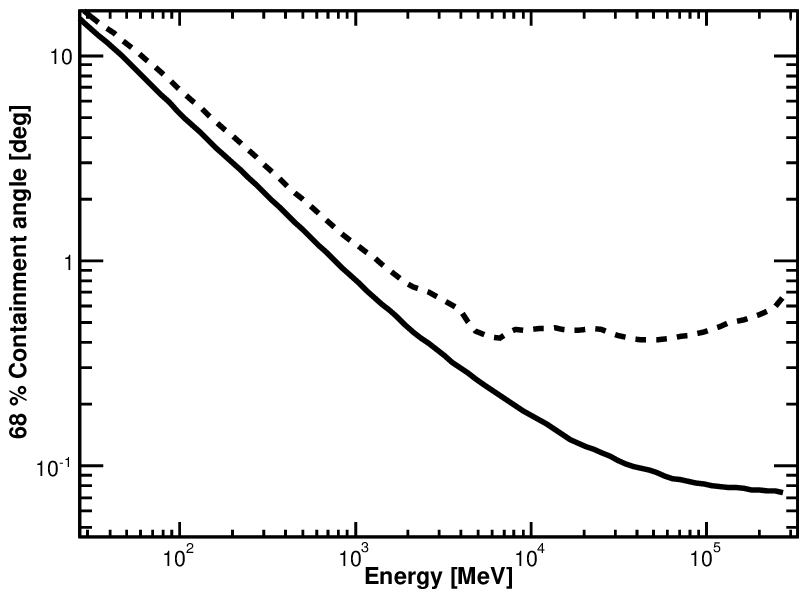}{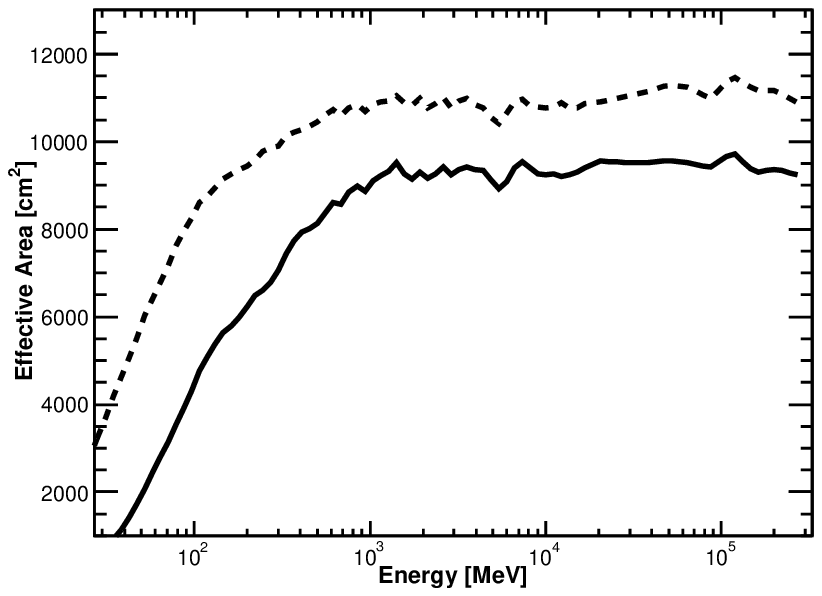}
  \caption{
    {\bf Left}: Comparison of the estimated Point Spread Function (PSF) for the
    onboard
    and on-ground event reconstruction and selection. The black solid curve is
    the 68\% containment angle on-axis for
    the transient event class, while the dashed curve represents the performance
    of the onboard reconstruction.
    {\bf Right}: Comparison of the estimated onboard (dashed) and on-ground
    (solid black curve) on-axis effective areas.  These estimates of the
    instrument response are based on simulations of the LAT.
  }
  \label{fig:irf_comparison}
\end{figure}

Key points of the LAT design are: wide Field-Of-View
(FOV---more than $2$~sr), large effective area and
excellent Point Spread Function (PSF---see
Fig.~\ref{fig:irf_comparison}), short dead time ($\sim 25$
$\mu$s per event) and good energy resolution (of the order
of 10\% in the central region of the active energy range).
As a result, the LAT is the most sensitive high energy
gamma-ray detector ever flown. The study of gamma-ray
bursts (GRBs) will take particular advantage of the
improvement in angular resolution---we estimate that two or
three photons above 1 GeV will localize a bursts to $\sim
5$~arcminutes. The reduced dead time will allow the study
of the sub-structure of the GRB pulses, typically of the
order of milliseconds \citep{Walker:00}, with a time
resolution that has never before been accessible at GeV
energies.

The data telemetered to the ground consists of the signals
from different parts of the LAT; from these signals the
ground software must `reconstruct' the events and filter
out events that are unlikely to be gamma-rays. Therefore,
the Instrument Response Functions (IRFs) depend not only on
the hardware but also on the reconstruction and event
selection software.  For the same set of reconstructed
events trade-offs in the event selection between retaining
gamma rays and rejecting background result in different
event classes.  There are currently three standard event
classes---the \emph{transient}, \emph{source} and
\emph{diffuse} event classes---that are appropriate for
different scientific analyses (as their names suggest).
Less severe cuts increase the photon signal (and hence the
effective area) at the expense of an increase in the
non-photon background and a degradation of the PSF and the
energy resolution.

The least restrictive class, the transient event class, is
designed for bright, transitory sources that are not
background-limited. We expect that the on-ground event rate
over the whole FOV above 100~MeV will be 2~Hz for the
transient class and 0.4~Hz for the source class. In both
cases we expect about one non-burst event per minute within
the area of the PSF around the burst position.
Consequently, there should be essentially no background
during the prompt emission (with a typical duration of less
than a minute) so that the transient class is the most
appropriate---and in fact is the one used for producing all
the results presented in this paper. On the other hand, the
analysis of afterglows, which may linger for a few hours,
will need to account for the non-burst background, at least
in the low region of the energy spectrum, where the PSF is
larger (see Fig.~\ref{fig:irf_comparison}).

The onboard flight software also performs event
reconstructions for the burst trigger.  Because of the
available computer resources, the onboard event selection
is not as discriminating as the on-ground event selection,
and therefore the onboard burst trigger is not as sensitive
because the astrophysical photons are diluted by a larger
background flux. Similarly, larger localization
uncertainties result from the larger onboard PSF, as shown
by the left-hand panel of Fig.~\ref{fig:irf_comparison}.

\subsection{$Fermi$ Gamma-ray Burst Monitor (GBM)}\label{sec:GBM_Description}

The GBM detects and localizes bursts, and extends $Fermi$'s burst
spectral sensitivity to the energy range between 8~keV and 30~MeV or
more. It consists of 12~NaI(Tl) (8--1000~keV) and 2~BGO
(0.15--$>30$~MeV) crystals read by photomultipliers, arrayed with
different orientations around the spacecraft. The GBM monitors more
than 8~sr of the sky, including the LAT's FOV, and localizes bursts
with an accuracy of $<15\arcdeg$ (1$\sigma$) onboard, ($<3\arcdeg$ on ground), 
by comparing the rates
in different detectors. The GBM is described in greater detail in 
Meegan et al. (2009, submitted).

\subsection{$Fermi$'s Burst Operations}\label{sec:Burst_Operations}

Both the GBM and the LAT have burst triggers. When either
instrument triggers, a notice is sent to the ground through
the TDRSS within $\sim 15$~s after the burst was detected
and then disseminated by the Gamma-ray burst Coordinates
Network (GCN)\footnote{See http://gcn.gsfc.nasa.gov/} to
observatories around the world. This initial notice is
followed by messages with localizations calculated by the
flight software of each detector. Additional data (e.g.,
burst and background rates) are also sent down by the GBM
through TDRSS for an improved rapid localization on the
ground by a dedicated processor.

Updated positions are calculated from the full datasets
from each detector that are downlinked with a latency of a
few hours.  Scientists from both instrument teams analyze
these data, and if warranted by the results, confer.
Conclusions from these analyses are disseminated through
GCN Circulars, free-format text that is e-mailed to
scientists who have subscribed to this service.  Both
Notices and Circulars are posted on the GCN website.

If the observed burst fluxes in either detector exceed
pre-set thresholds (which are higher for bursts detected by
the GBM outside the LAT's FOV), the FSW sends a request that
the spacecraft slew to point the LAT at the burst location
for a followup pointed observation; currently a 5~hr
observation is planned.  

In addition to the search for GRB onboard the LAT and  manual follow-up analysis by duty scientists, there is also automated processing of the full science data. This processing performs an independent search for transient events in the LAT data, to greater sensitivity than is possible onboard, and also performs a counterpart search for all GRB detected within the LAT FoV. This is described in greater detail in \S~\ref{sec:LAT_Ground}.

%

%
%

\section{Burst Simulations}\label{sec:Burst_Simulations}

We test the $Fermi$ burst detection and analysis software with
simulated data.  These simulated data are based on our expectations
for burst emission in the LAT and GBM spectral bands (see
\S~\ref{sec:Expectations}), and on models of the instrument response
of these two detectors. Since bursts undoubtedly differ from our
theoretical expectations, our calculations are more reliable in
showing the mission's sensitivity to specific bursts than in
estimating the number of bursts that will be detected.

We have two `GRB simulators' that model the burst flux incident on
each detector \citep{Battelino:07a}.  The primary is the
phenomenological simulator---described in greater detail below in
\S~\ref{sec:Phenomenological}---that draws burst parameters from
observed distributions. We have also created a physical simulator
\citep{Omodei:05, Omodei:07, Omodei:07b}
that calculates the synchrotron emission from the collision of
shells in a relativistic outflow (the internal shock
model---\citealt{Piran:99}).
For a given analysis we assemble an ensemble of simulated
bursts using one of these GRB simulators.  To simulate a
LAT observation of each burst in this ensemble we create a
realization of the photon flux, resulting in a list of
simulated photons incident on the LAT. The LAT's response
to this photon flux is processed in one of two software
paths. The first uses `GLEAM', which performs a Monte Carlo
simulation of the propagation of the photon and its
resulting particle shower in the LAT (using the GEANT4
toolkit\citep{G4:2003}) and the detection of particles in the different
LAT components\citep{Atwood:2004,Baldini:06}. The photon is then `reconstructed' from
this simulated instrument response by the same software
that processes real data.  Thus GLEAM maps the incident
photons into observed events. Our second, faster,
processing pathway uses the instrument response functions
to map the photons into events directly. We note that both
approaches use the same input---a list of incident
photons--and result in the same output---a list of
`observed' events in one of the event classes. In both
approaches GRBs can be combined with other source types
(such as stationary and flaring AGN, solar flares,
supernova remnants, pulsars) to build a very complex model
of the gamma-ray sky.

The GRB simulators also provide the input to the GBM simulation
software.  In this case the GRB simulators produce a time series of
spectral parameters (usually the parameters for the `Band'
function---\citealt{Band:03}---discussed above in
\S~\ref{sec:Previous_Observations}). The GBM simulation software
samples the burst spectrum to create a list of incident photons and
then uses a model of the GBM response to determine whether each
photon is `detected,' and if so, in which energy channel (simulating
the GBM's finite spectral resolution). Based on a model from the
BATSE observations, background counts are added to the burst counts.
The GBM simulation software outputs count lists, response matrices
and background spectra in the standard FITS formats used by software
such as XSPEC.\footnote{See http://heasarc.nasa.gov/xanadu/xspec/}

\begin{figure}
  \plotone{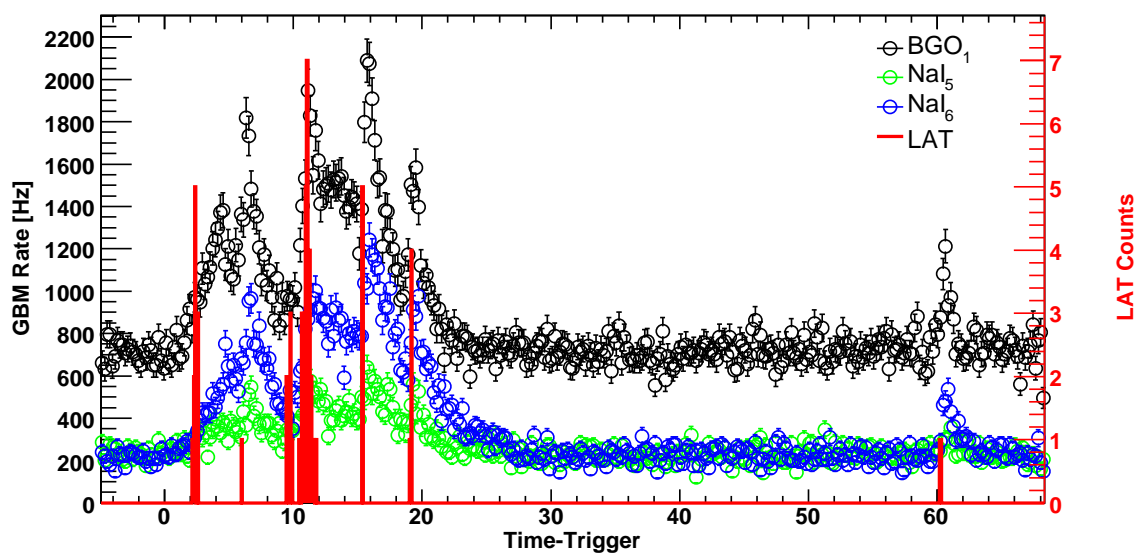}
  \caption{Simulated count rate light curve for a BGO detector, two NaI detectors,
and the LAT for one simulated burst. In this model of the burst
spectral evolution, the LAT detects counts at the beginning of
each pulse; the correlation of the LAT and GBM light curves will be
a powerful diagnostic of the emission processes. {\nicola The
simulation predicts that the LAT would detect a total of 42 gamma rays
above 30 MeV in this moderately bright burst of 1~s peak
flux of 63.37 ph cm$^{-2}$ s$^{-1}$ between 30 and 500 keV.}}
   \label{fig:LC}
\end{figure}

Because the GRB simulators provide input to both LAT and GBM
simulations, simulated LAT and GBM data can be produced for the same
bursts, allowing joint analyses.  The $Fermi$ mission developed the
`Standard Analysis Environment' (SAE) to analyze both LAT and GBM
data. Data can be binned in time, resulting in light curves (see,
for example, Fig.~\ref{fig:LC}), or in spectra that can be analyzed
using a tool such as XSPEC.  As will be described in
\S~\ref{sec:Spectral_Analysis}, joint fits of GBM and LAT data may
cover an energy band larger than seven orders of magnitude (see
Fig.~\ref{fig:SP_1}).  Consequently, $Fermi$ will be a very powerful
tool for understanding the correlation between low-energy and
high-energy GRB spectra.

\subsection{Phenomenological Burst
Model}\label{sec:Phenomenological}

The phenomenological GRB simulator that is used for most of our
simulations draws from observed spectral and temporal distributions
to construct model gamma-ray bursts.  This modeling assumes that
bursts consist of a series of pulses that can be described by a
universal family of functions \citep{Norris:96}
\begin{equation}
\label{eqn:pulse}
I(t) = A \left\{
   \begin{array}{ll}
      \exp[-(|t-t_0 |/\sigma_r)^\nu],  \hspace{1cm} t \leq t_0\\
\\
      \exp[-(|t-t_0 |/\sigma_d)^\nu], \hspace{1cm} t>t_0\\
   \end{array}
\right.
\end{equation}
where $\sigma_r$ and $\sigma_d$ parameterize the rise and decay
timescale, and $\nu$ provides the `peakiness' of the pulse.
Although empirically $\sigma_r\sim0.33~\sigma_d^{0.86}$, we
approximate this relation as $\sigma_r\sim \sigma_d/3$. The pulse
Full Width at Half Maximum (FWHM) is
\begin{equation}
\label{eqn:W}
W=(\sigma_r+\sigma_d)\ln(2)^{1/\nu}.
\end{equation}
Pulses are observed to narrow at higher energy in the BATSE energy
band \citep{Davis:94,Norris:96,Fenimore:95}. Although the statistics
in the EGRET data were insufficient to determine whether this
narrowing continues in the $>$100~MeV band, our phenomenological
model assumes that it does.  We assume that the FWHM energy
dependence is $W(E)\propto E^{-\xi}$ where $\xi$ is $\sim$0.4
\citep{Fenimore:95,Norris:96}. Thus, we give the pulse shape in
eq.~\ref{eqn:pulse} an energy dependence by setting
\begin{equation}
\label{eqn:srsd_e}
\left\{
   \begin{array}{ll}
       \sigma_d(E) = 0.75\times \ln(2)^{-1/\nu} W_0 (E/20 \hbox{ keV})^{-\xi}\\
\\
       \sigma_r(E)=0.25\times \ln(2)^{-1/\nu} W_0 (E/20 \hbox{ keV})^{-\xi}.
   \end{array}
\right.,
\end{equation}
where $W_0$ is the FWHM at 20~keV.  Burst spectra in the
10--1000~keV band are well-described by the `Band' function
\citep{Band:93} parameterized in eq.~\ref{eqn:BandF2}.  Empirically
the Band function is an adequate description of burst spectra
accumulated on short timescales (e.g., shorter than a pulse width)
and over an entire burst.  This may be due in part to the poor
spectral resolution of scintillation detectors (such as BATSE and
the GBM), but we will treat this as a physical characteristic of
gamma-ray bursts.  In the resulting model, the flux $f(t,E)$ is a
product of a Band function with spectral indices $\alpha^\prime$ and
$\beta^\prime$ and the energy-dependent pulse shape $I(t,E)$
(eq.~\ref{eqn:pulse} with eq.~\ref{eqn:srsd_e})
\begin{equation}
\label{eqn:model1}
f(t,E)=I(t,E)~N_{\rm Band}(E | N_0, E_p,\alpha^\prime,\beta^\prime) \hspace{1cm}
   \hbox{ph cm$^{-2}$ s$^{-1}$ keV$^{-1}$} .
\end{equation}
Note that this spectrum is not strictly a Band function because the
pulse shape function does not have a power law energy dependence.

The spectrum integrated over the entire burst is a Band function
that is proportional to the product $W(E) N_{\rm Band}(E | N_0, E_p,
\alpha', \beta')$. Because $W(E)$ is a power law with spectral index
-$\xi$, the spectral indices $\alpha$ and $\beta$ for the integrated
spectrum are different from the indices for the instantaneous flux
(eq.~\ref{eqn:model1})
\begin{equation}
\label{eqn:model5}
\begin{array}{ll}
    \Int_{-\infty}^{\infty} f(t,E) dt &=~ N_{\rm Band}(E | N_0,E_p,\alpha,\beta) T =
    A_0~ N_{\rm Band}(E | N_0,E_p,\alpha^\prime,\beta^\prime)~ W(E) \\
&\\
& =~ A_0~W_0~N_{\rm Band}(E | N_0,E_p,\alpha^\prime-\xi,\beta^\prime-\xi)
\end{array}
\end{equation}
where $T$ is the burst duration and all the normalizing factors
resulting from the integration are incorporated in $A_0$.  Thus the
flux for a single GRB is the sum of many pulses of the form
\begin{equation}
\label{eqn:model6}
f(t,E) =  I(t,E) N_{\rm Band}(E | N_0,E_p,\alpha+\xi,\beta+\xi) .
\end{equation}
Drawn from observed burst distributions, the same spectral
parameters $E_p$, $\alpha$ and $\beta$ are used for a given
simulated burst.  The number of pulses and parameters of each pulse
(amplitude, width and peakedness) are also sampled from observed
distributions \citep{Norris:96}.

Alternative spectral models have also been simulated; for example,
\citet{Battelino:07a} describe simulations with a strong thermal
photospheric component.

%

%
%

\section{Semi-Analytical Sensitivity Estimates}\label{sec:Design_Sensitivity}

The design of the LAT detector provides an ultimate burst
sensitivity, regardless of whether the detection and
analysis software achieves this ultimate limit.  Thus in
this section we estimate the LAT's burst detection and
localization capabilities, and the expected flux
distribution.  The following section describes the current
burst detection algorithms.

\subsection{Semi-Analytical Estimation of the Burst
Detection
Sensitivity}\label{sec:Semi_Analytical_Sensitivity}

In this subsection we compute the LAT's burst detection sensitivity
using a semi-analytical approach based on the likelihood ratio test
introduced by \citet{Neyman:28}. This test is applied extensively to
photon-counting experiments \citep{Cash:79} and has been used to
analyze the gamma-ray data from COS-B \citep{Pollock:81,Pollock:85}
and EGRET \citep{Mattox:96}. The statistic for this test is the
likelihood for the null hypothesis for the data divided by the
likelihood for the alternative hypothesis, here that burst flux is
present. This methodology is the basis of the likelihood tool that
will be used to analyze LAT observations; here we perform a
semi-analytic calculation for the simple case of a point source on a
uniform background.

%
%

In photon-counting experiments, the natural logarithm of the
likelihood for a given model can be written as
\begin{equation}
   \ln(L) =\sum_{photons} \ln(M_i) - N_\mathrm{pred} + \mathrm{constant}
   \label{eq:ln_L}
\end{equation}
where $M_i$ is the predicted photon density at the position and time
of $i$th observed count, and $N_\mathrm{pred}$ is the predicted
total number of counts.  We compare the log likelihood for the null
hypothesis that only background counts are present versus the
hypothesis that both burst and background counts are present.

The expected number of counts from a burst flux $S(E)$ is
\begin{equation}
N_S =  T_{obs}\int_{\Delta\Omega}\int_{E_1}^{E_2}
   A_{eff}(E) S(E)F(E,\Omega) \,dE d\Omega
\label{eq:N_S}
\end{equation}
while the expected number of counts from a background flux
$B(E)$ (assumed to be uniformly distributed over the sky)
is
\begin{equation}
N_{B} =  T_{obs}\int_{E_1}^{E_2} A_{eff}(E)  B(E) dE \Delta\Omega
\label{eq:N_B}
\end{equation}
where $A_{eff}$ is the effective area and $F(E,\Omega)$ is
the normalized PSF (which therefore does not show up in
eq.~\ref{eq:N_B}).  Note that $B(E)$ varies significantly
over the sky, but our assumption is that it is constant
over $\Delta\Omega$.

The logarithm of the likelihood of the null hypothesis is
\begin{eqnarray}
\ln(L_0) = T_{obs}\int_{\Delta\Omega}\int_{E_1}^{E_2} A_{eff}(E)
   \left[S(E)F(E,\Omega) + B(E)\right] \times \nonumber \\
   \ln(A_{eff}(E) B(E))dE d\Omega  - N_{B} \quad .
   \label{eq:ln_L0}
\end{eqnarray}
The actual count rate is assumed to result from both
background and burst flux while the predicted count rates
(the $M_i$ in eq.~\ref{eq:ln_L} and the total number of
counts $N_{\rm pred}$) are calculated only for the
background flux (the null hypothesis).

Similarly, the logarithm of the likelihood of the
hypothesis that a burst is present is
\begin{eqnarray}
\ln(L_1) = \left[T_{obs}\int_{\Delta\Omega}\int_{E_1}^{E_2}
   A_{eff}(E) \left[S(E)F(E,\Omega) + B(E)\right] \right. \times \nonumber \\
   \left.\ln\left(A_{eff}(E) \left[S(E)F(E,\Omega) +
   B(E)\right]\right) dE d\Omega\right]  - (N_S+N_B) \quad .
   \label{eq:ln_L1}
\end{eqnarray}
Here both the actual and predicted count rates are calculated for
both burst and background fluxes.

Wilks' theorem \citep{Wilks:38} defines the Test Statistic
as $T_{S}=-2(\ln(L_0)-\ln(L_1))$, and states that $T_S$ is
distributed (asymptotically) as a $\chi^2$ distribution of
$m$ degrees of freedom, where $m$ is the number of burst
parameters. From eqs.~\ref{eq:ln_L0} and \ref{eq:ln_L1}
$T_{S}$ is
\begin{equation}
T_{S} =2~T_{obs}\int_{\Delta\Omega}\int_{E_1}^{E_2} A_{eff}(E) B(E)
   \left[ \left(1+G(E,\Omega)\right)
   \ln\left(1+G(E,\Omega)\right) -G(E,\Omega) \right] dE d\Omega
\label{eq:Ts}
\end{equation}
where we have defined a signal-to-noise ratio
$G(E,\Omega)=S(E)F(E,\Omega)/B(E)$.

The significance of a source detection in standard
deviation units is calculated as $N_\sigma=\sqrt{T_{S}}$ in
the case $m=1$ ($\chi^2$ with 1 dof). Here we assume that
Wilks' theorem holds, which might be not absolutely true in
a low-count regime (see, in particular, the discussion in
\S~\ref{sec:UL}). However, we will see that this method
gives a robust estimate of the LAT sensitivity to GRBs. We
can use this method to estimate the LAT sensitivity to GRB.

\begin{figure}
  \epsscale{0.5}
  \plotone{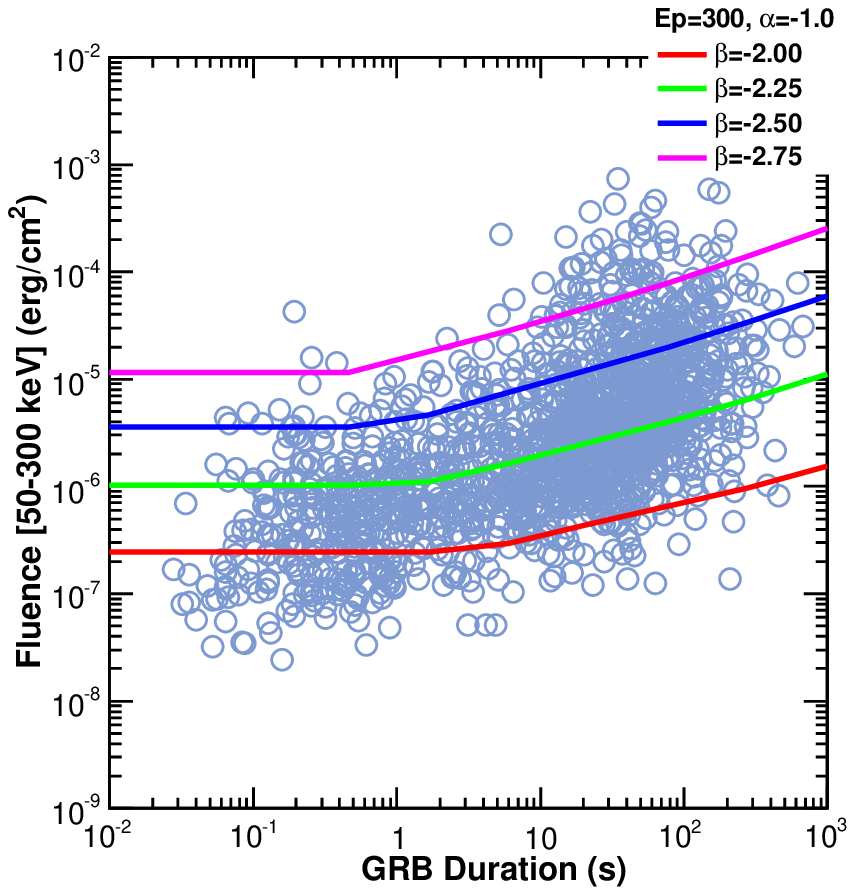}
  \caption{{\nicola Threshold fluence as a function of the GRB duration, for
  on-ground detection and for on-axis incidence. 
  Threshold fluence increases by factor of $\sim$ 2 for z-axis angles of 50 degrees.
  Different lines are related to different spectral index.} Also plotted
  are the observed bursts from the BATSE catalog.}
  \epsscale{1}
  \label{fig:sensitivity_FluenceVsDuration}
\end{figure}

In our modeling we assume the burst has a `Band' function spectrum
(see eq.~\ref{eqn:BandF2}) and that the flux is constant over a
duration $T_{GRB}$.  Since we seek the optimal detection
sensitivity, we calculate $T_S$ for $T_{obs}=T_{GRB}$.  We assume a
spatially uniform background with a power law spectrum
\begin{equation}
B(E)=B_{0}\left(\frac{E}{\hbox{100 MeV}}\right)^{\gamma}
   \quad \hbox{ph
cm$^{-2}$ MeV$^{-1}$ s$^{-1}$ sr$^{-1}$}
 \label{be}
\end{equation}
where the value of the normalization constant $B_{0}$ is
set to mimic the expected background rate. For modeling the
onboard trigger the background rate {\nicola above 100 MeV} is set to 120~Hz,
while, for the on-ground trigger the background is set to
2~Hz, as will be discussed below. The spectral index is set
to be $\gamma=-2.1$. The results depend on the value of the
spectral index; a detailed study of the dependence of the
results as a function of the shape of the residual
background is outside the illustrative goal of this
section, thus we omit such discussion. We require
$T_{S}\ge25$ and at least 10 source counts in the LAT
detector, corresponding to a threshold significance of
5$\sigma$ and a minimum number of GRB counts {\nicola to
see a clear excess in the LAT data even in the case of very
few background events}. We use the ``transient'' event
class described in \S~\ref{sec:LAT_Description}, and
compute the minimum 50--300~keV fluence of bursts at this
detection threshold. The burst fluxes in the LAT band
depend only on the high energy power law component of the
`Band' spectrum; assumed values of the low energy power law
spectral index $\alpha=-1$ and $E_p=500$~keV are used to
express the spectrum's normalization in familiar fluence
units. Results are shown in
Fig.~\ref{fig:sensitivity_FluenceVsDuration}; at short
durations the threshold is determined by the finite number
of burst photons, while the background determines the
threshold for longer durations.  This figure predicts that
unless other high-energy spectral components are present,
the bursts detected by the LAT will be `hard' with photon
indices $\beta$ near $-2$ \citep{Band:07}.

\begin{figure}
  \plottwo{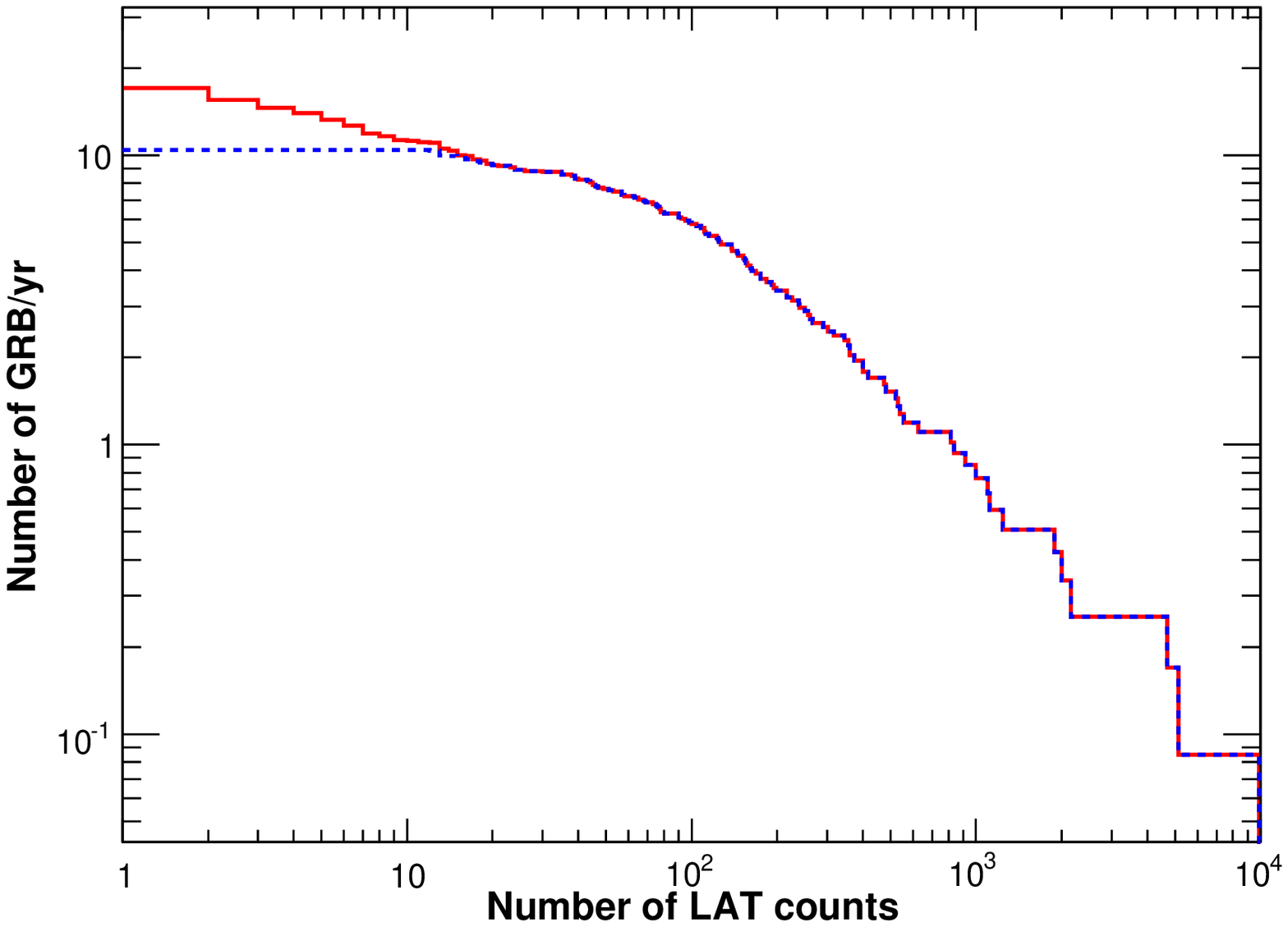}
          {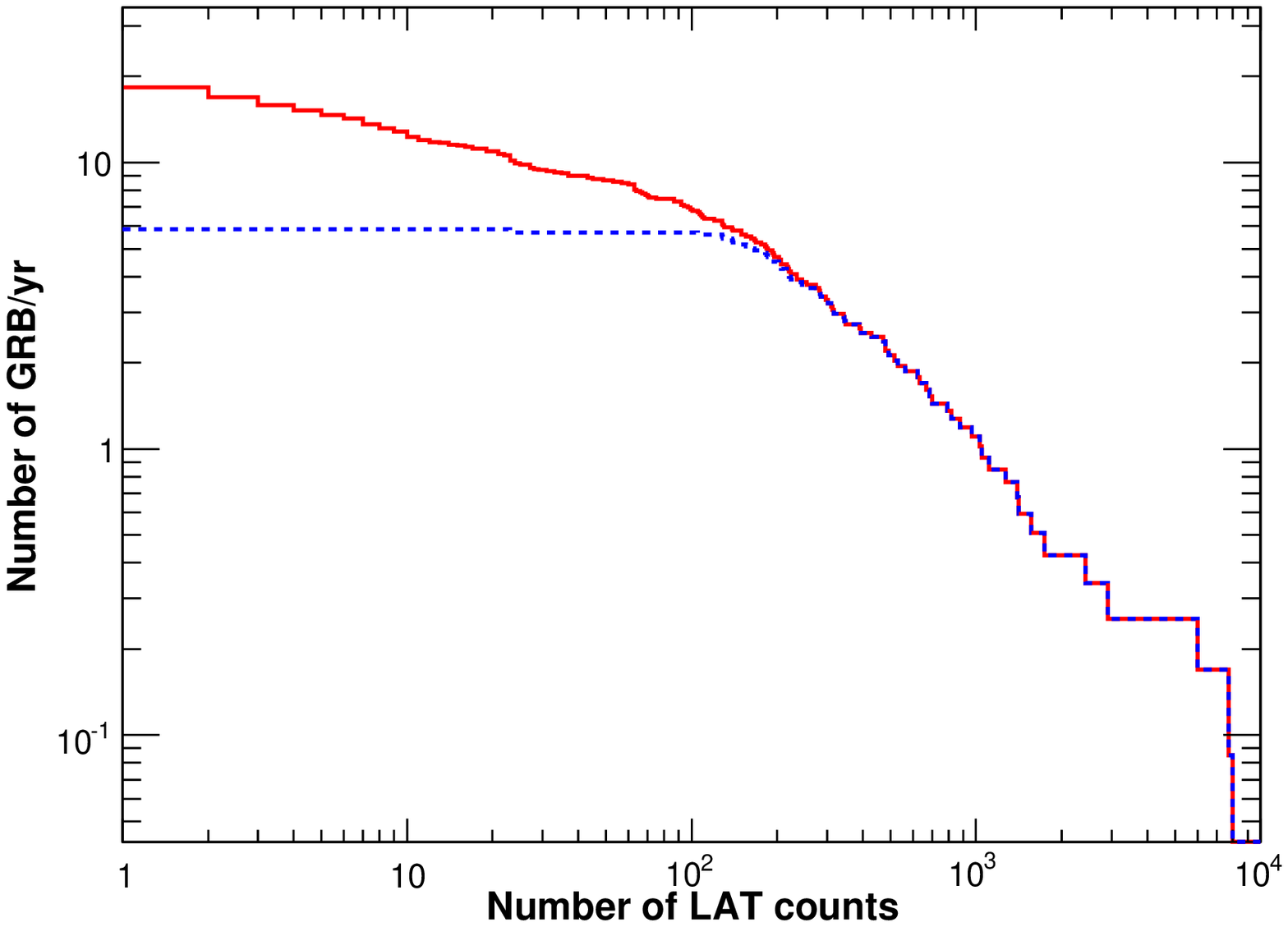}
  \caption{Integrated number of GRBs per year as a function of the
  number of LAT counts.  The solid curve shows all bursts in the
  sample, while the dashed curve gives the detected bursts.  Left
  panel: on-ground analysis (``transient'' class, 2~Hz background
  rate above 100~MeV).  Right panel: onboard analysis (120~Hz background rate).
  }
  \label{fig:sensitivity_like}
\end{figure}

These estimates consider the detectability of individual bursts.  
{\nicola
We can compute the sensitivity of the LAT detector to GRB considering as input the observed distribution of GRB with known spectral parameters. 
We use the catalog of bright bursts \citep{Kaneko:06} to quantify the characteristics of GRBs. This catalog contains 350 bright GRBs over the entire life of the BATSE experiment selected for their energy fluence (requiring that the fluence in the 20-2000~keV band is greater than 2$\times$10$^{-5}$~erg/cm$^2$) or on their peak photon flux (over 256 ms, in the 50-300~keV, greater than 10~ph/cm$^2$/s). This subset of burst of the whole BATSE catalog represents the most comprehensive study of spectral properties of GRB prompt emission to date and is available electronically from the High-Energy Astrophysics Science Archive Research Center (HEASARC)\footnote{http://heasarc.gsfc.nasa.gov/}. We restrict our sample of GRB to the ones with a well reconstructed $E_{peak}$; furthermore, we exclude the bursts described by the Comptonized model (COMP) for which an emission at LAT energy is very unlikely; 
we also reject bursts with spectra described by a single power law with undetermined $E_{peak}$ (probably outside the BATSE energy range).

Considering the field of view of the BATSE experiment and these selection criteria, we estimate a rate of 50 GRB per year (full sky).
For each burst we simulate, the duration, the energy fluence and the spectral parameters are in agreement with one of the bursts in the Bright BATSE catalog. Its direction is randomly chosen in the sky, and for each burst we compute the LAT response functions for that particular direction. Finally, we compute $T_s$ using eq.~\ref{eq:Ts}. The resulting distributions are given by
Fig.~\ref{fig:sensitivity_like}.
}


%

The onboard analysis' larger effective area
(Fig.~\ref{fig:irf_comparison}) results in a larger
cumulative burst rate, but not a larger detected rate
because of the larger background rate.  {\nicola Events
that are processed onboard by the GRB search algorithm are downloaded, 
and a looser set of cuts can be chosen
on-ground in order to optimize the signal/noise ratio}. We emphasize that this calculation makes a number of
simplifying assumptions. The LAT spectrum is assumed to be
a simple extrapolation of the spectrum observed by BATSE.
Spectral evolution within a burst is not considered.  The
BATSE burst population was biased by that instrument's
detection characteristics. Nonetheless we estimate that the
LAT can detect around {\nicola 1 burst per month, with a few bursts per year having more than 100 counts. These few bright bursts are
likely to have a large impact on burst science since
detailed spectral analysis will be possible.}

\begin{figure}
   \plottwo{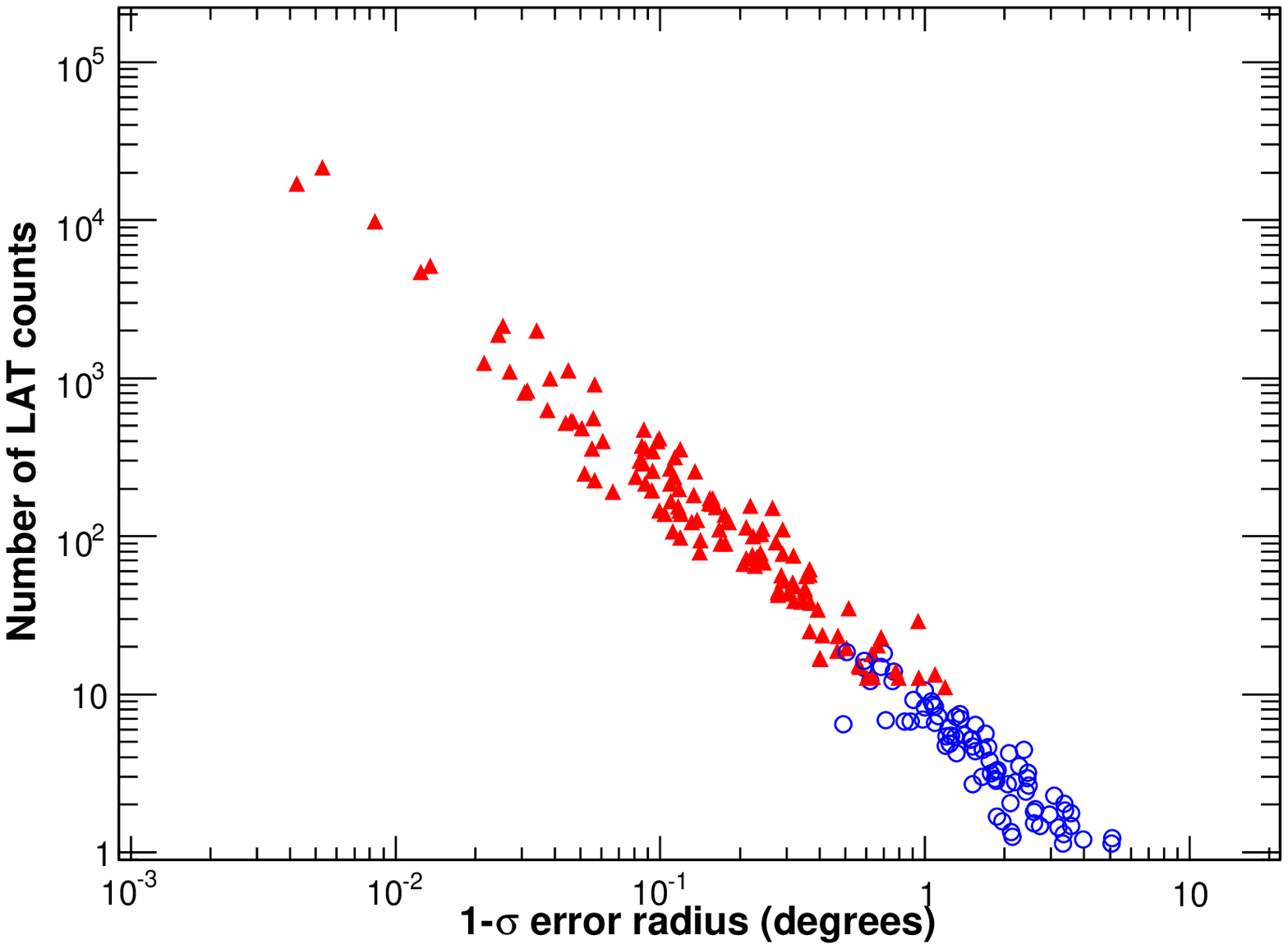}
	   {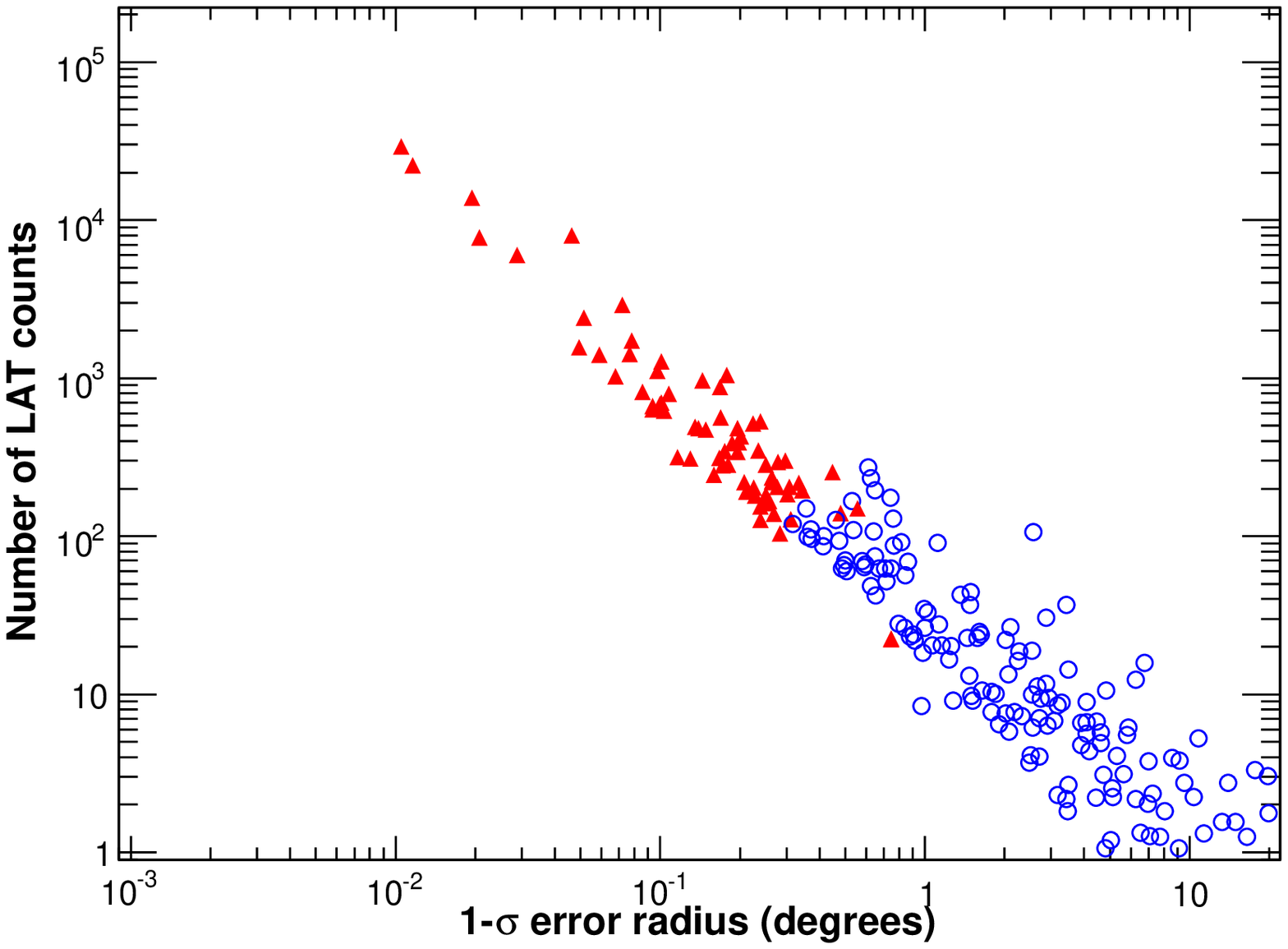}
       \caption{Number of LAT counts vs. localization accuracy.
       In each panel the red triangles denote detected bursts and the
       open blue circles show undetected bursts.  The left and right
       panels are for the on-ground and onboard localizations.  Thus
       the on-ground analysis results in a {\nicola slightly} larger
       burst detection
       rate and a better localizations.  The superior track reconstruction
       and background reduction outweighs the smaller effective
       area in increasing the on-ground detection rate.}
   \label{fig:localization_like}
\end{figure}

In the framework described in this section, we can also
estimate the localization accuracy for the burst sample,
for both onboard and on-ground triggers. If $\sigma_i$ is
the 68$\%$ containment radius for the single photon PSF,
then the localization is computed as
\begin{equation}
\sigma_{GRB}^{-1}=\sqrt{\sum_i{\frac{1}{\sigma_i^{2}}}}
\end{equation}
that, in terms of the previously defined quantities, is
\begin{equation}
\sigma_{GRB}^{-1}=\sqrt{\frac{T_{GRB}}{3} \int_{E_{1}}^{E_{2}}
   \frac{ A_{eff}(E) S(E)}{\sigma_{68\%}(E)^{2}} dE}
\end{equation}
The factor of 3 takes into account the non-gaussianity of
the PSF, and was estimated by \cite{Burnett:07}. We compute
the localization accuracy for each burst in our sample.
Fig.~\ref{fig:localization_like} shows the results. In each
plot the detected burst are represented by red triangles,
while the blue empty circles are the bursts with LAT counts
that did not pass our detection condition.

These results show that the LAT can localize bursts with sub-degree
accuracy, both onboard and on-ground. The GRB yield is greater and
bursts are better-localized on-ground than onboard. The on-ground
analysis is available only after the full dataset is downlinked and
processed. This process can lasts few hours, depending on the
position of the downlink contact. Onboard localization is delivered
quasi-real time with onboard alerts. For those bursts,
multiwavelength follow-ups will be feasible for bursts localized
within a few tens of arcminutes. For example, the FOV of {\it
Swift's} XRT is about 0.4$^\circ$ and is of the same order as the
FOV of the typical mid-size optical or near-IR (NIR) telescope.
Afterglow searches in the optical and NIR are very
successful---$\sim$60\% of the {\it Swift} bursts have been
associated with optical and NIR afterglows.
Fig.~\ref{fig:localization_like} shows that a sizeable fraction of
$Fermi$ GRB detections will be localized within these requirements,
and relatively large FOV ground-based observatories ($\sim$30
arcmin) with optical/NIR filters (I, z, J, H, K) should produce a fairly
high detection rate for the afterglows of LAT-detected GRBs.

%
%

\subsection{Estimated LAT Flux Distribution}
\label{sec:Sensitivity}

We now consider the full GRB model described in
\S~\ref{sec:Burst_Simulations} for estimating the expected
LAT flux distribution. This is, of course, very dependent
on the assumptions of the GRB model, and the final result
should be considered only as a prediction of the flux
distribution. 

{\nicola We use the bright BATSE catalog \citep{Kaneko:06} for the burst
population, as described in the previous section.
In addition, we also select a sub-sample of bursts for which beta is more negative than -2. This is motivated by the fact that a power law index greater than -2 implies a divergence in the released content of energy, thus those value are unphysical and a cut-off should take place.
The measurements yielding beta greater than -2 are questionable and suggest either an ill-determined quantity for a true spectrum that is in reality softer, or an additional spectral break above the energies measured with BATSE. 
Given the duration, the number of pulses is fixed by the total burst duration. 
Pulses are combined together in order to obtain a final $T_{90}$ duration. 
Correlations between duration, intensity, and spectral parameters are automatically taken into account as each of these bursts corresponds to an entry in the Kaneko et al. catalog.
The emission is extended up to high energy with the model described in
\S~\ref{sec:Burst_Simulations}. 

We emphasize again that
this model ignores possible intrinsic cutoffs (resulting
from the high end of the particle distribution or internal
opacity---\S~\ref{sec:HE_abs}), and additional high-energy
components suggested by the EGRET observations
(\S~\ref{sec:Previous_Observations}). 
}
High-energy emission ($>$10~GeV)  is also sensitive to
cosmological attenuation due to pair production between the
GRB radiation and the Extragalactic Background Light
(EBL---\S~\ref{sec:HE_abs}). The uncertain EBL spectral
energy distribution resulting from the absence of high
redshift data provides a variety of theoretical models for
such diffuse radiation. Thus the observation of the
high-energy cut-off as a function of the GRB distance can,
in principle, constrain the background light. In our
simulation we include this effect, adopting the EBL model
in \citet{Kneiske:04}.  Short bursts are thought to be the
result of the merging of compact objects in binary systems,
so we adopt the short burst redshift distribution from
\citet{Guetta:05}, while long bursts are related to the
explosive end of massive stars, whose distributions are
well traced by the Star Formation History
\citep{Porciani:01}. 

\begin{figure}
   \plotone{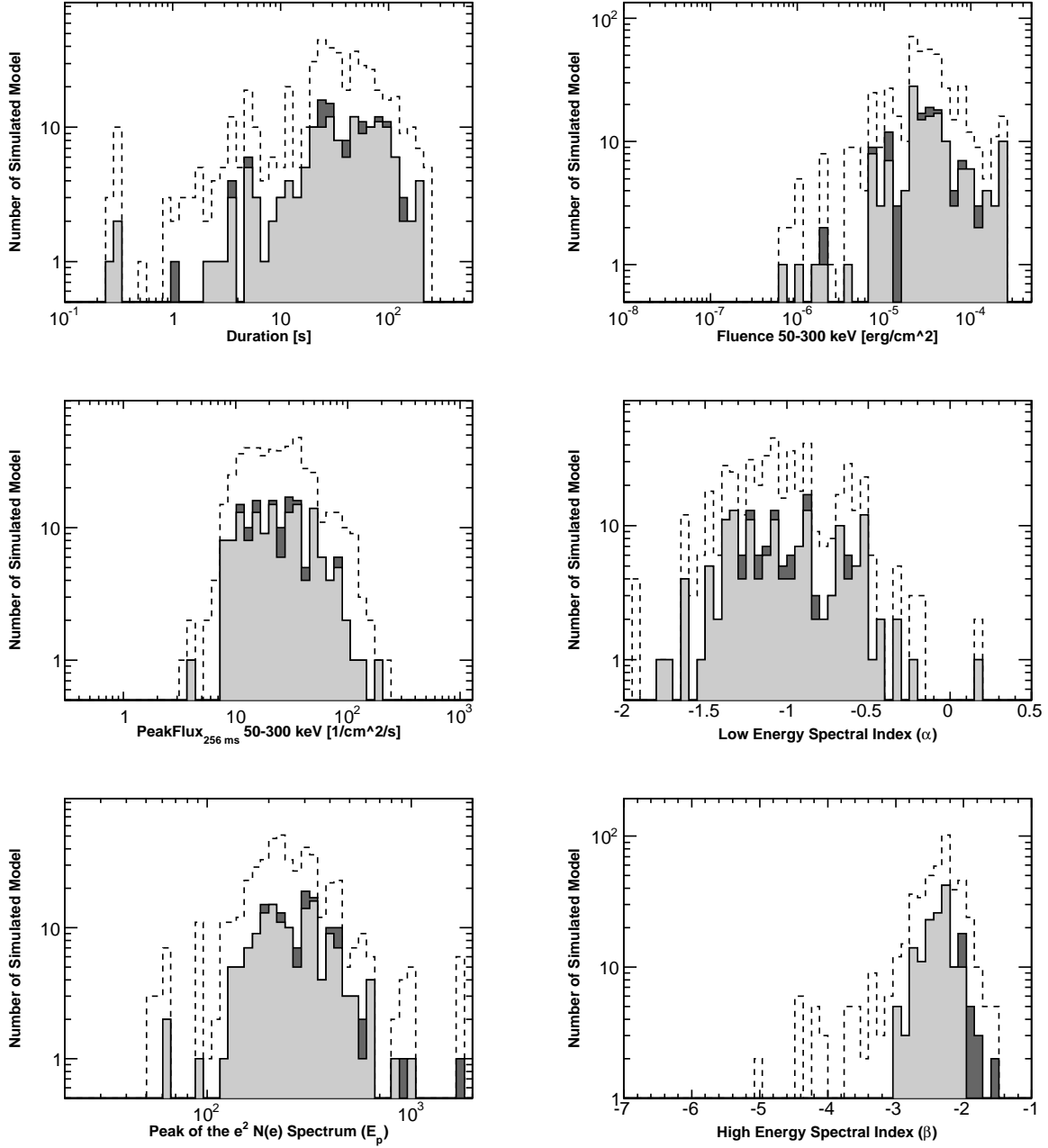}
   \caption{\nicola{Parameter distributions for the simulated bursts of the bright burst BATSE catalog (dashed lines). Filled dark histograms represent the GRBs with more than 1 predicted count above 100~MeV in the LAT detector, while for the light filled histograms we have also required that the high-energy spectral index beta is more negative than -2. The distributions show
the logarithm of the duration, the fluence, the peak flux distribution, 
the low and high energy spectral indexes and the
logarithm of the energy of the peak of the $\nu F_{\nu}$ spectrum.}}
   \label{fig:distributions}
\end{figure}

{\nicola In Fig.~\ref{fig:distributions} the
sampled distributions are shown. The Dashed line histogram is obtained from the full bright burst BATSE catalog. In order to increase the number of burst in the field of view of the LAT detector we over-sampled the original catalog by a factor 1.4.
The dark filled histograms show the distribution of GRB with at least 1 count in the LAT detector, and the light filled histograms are the sub-sample of detected GRB with beta $<$~-2.}

\begin{figure}
   \plotone{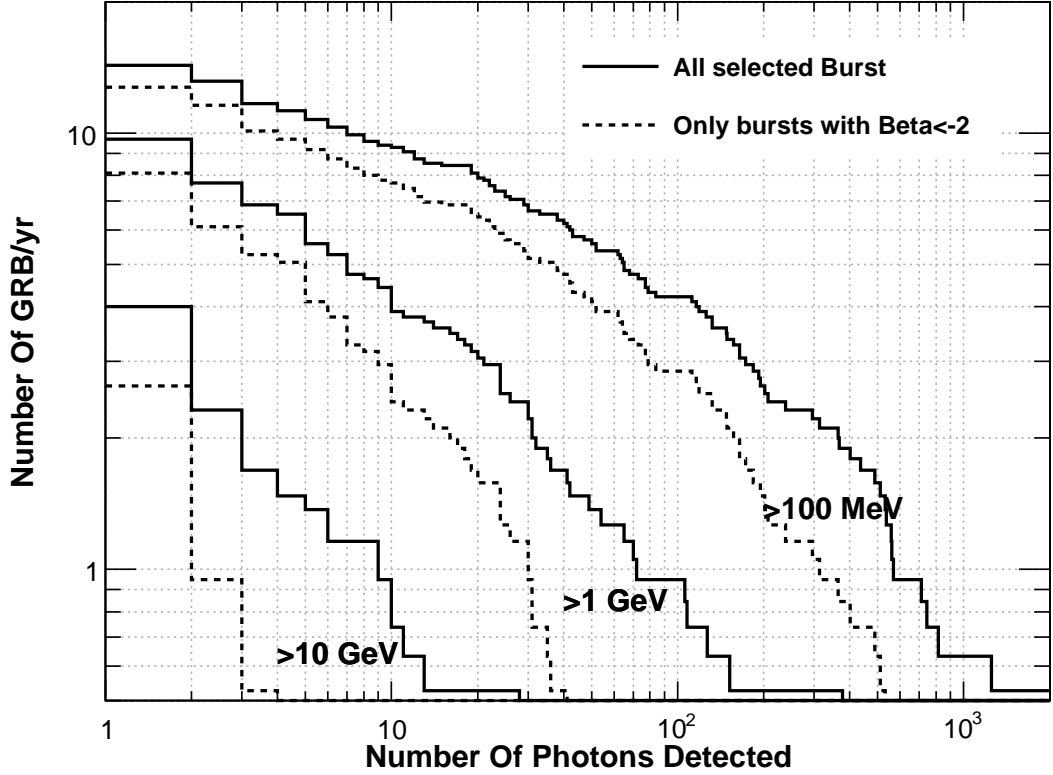}
   \caption{Model-dependent LAT GRB sensitivity. The GRB spectrum
is extrapolated from BATSE to LAT energies. The all-sky burst rate
is assumed to be 50 GRB yr$^{-1}$ full sky (above the peak flux in 256~ms of 10~ph~s$^{-1}$~cm$^{-2}$ in the 50-300~keV or with an energy flux in the 20-2000~keV band greater than 2$\times$ 10$^{-5}$~erg/cm$^2$), based on BATSE catalog of bright bursts. 
The effect of the EBL absorption is included. Different curves refer to
different energy thresholds. Dashed curves are the result of the analysis excluding very hard bursts, with a beta greater than -2.}
   \label{fig:sensitivity}
\end{figure}

We simulate {\nicola approximately} ten years of observations in scanning mode. The orbit of
the $Fermi$ satellite, the South Atlantic Anomaly (SAA) passages and
Earth occultations are all considered.
{\nicola
 In Fig.~\ref{fig:sensitivity}
we plot the number of expected bursts per year as a function of the
number of photons per burst detected by the LAT. The different couples of lines refer to different energy thresholds (100~MeV, 1~GeV, and 10~GeV). 
Dashed lines are the same computation but using only the sub-sample of GRBs with beta more negative than -2 (the light filled distribution in Fig.~\ref{fig:distributions}).
}
The EBL attenuation affects only the high-energy curve, as expected from the theory,
leaving the sensitivities almost unchanged below 10~GeV. Assuming
that the emission component observed in the 10--1000~MeV band
continues unbroken into the LAT energy band, we estimate that the
LAT will independently detect {\nicola approximately 10 bursts per year}, depending on the
sensitivity of the detection algorithm; {\nicola approximately one
burst every three months will have more than a hundred counts in the LAT
detector above 100~MeV: these are the bursts for which a detailed spectral or even
time resolved spectral analysis will be possible.
If we restrict our analysis to the sub-sample of bursts with beta more negative than -2, these numbers decrease. Nevertheless, even if we adopt this conservative approach, LAT should be able to detect independently approximately 1 burst every two months, and will be able to detect radiation up to tens of GeV.

With the assumed high-energy emission model a few bursts per year will show
high-energy prompt emission, with photons above 10~GeV.  These rates are in agreement with the
number of bursts detected in the LAT data after few months (GRB080825C \citep{Bouvier:08}, GRB080916C \citep{Tajima:08}, GRB081024B \citep{Omodei:08}), but the statistics is still low for any strong constraint on the burst population.}


%

\section{Gamma-Ray Burst Detection}\label{sec:GRB_Detection}

The rapid detection and localization of bursts is a major goal of
the $Fermi$ mission.  Both $Fermi$ instruments will search for bursts
both onboard and on-ground. These searches will detect bursts on
different timescales and with different sensitivities. Here we focus
on LAT burst detection, but for completeness we describe briefly GBM
burst detection.

\subsection{GBM Burst Detection}\label{sec:GBM_Detection}

Onboard the $Fermi$ observatory the GBM will use rate triggers that
monitor the count rate from each detector for a statistically
significant increase. Similar to the BATSE detectors, the GBM as a
whole will trigger when two or more detectors trigger.  A rate
trigger compares the number of counts in an energy band $\Delta E$
over a time bin $\Delta t$ to the expected number of background
counts in this $\Delta E$--$\Delta t$ bin; the background is
estimated from the rate before the time bin being tested.  The GBM
trigger uses the twelve NaI detectors with various energy bands,
including $\Delta E$=50--300~keV, and time bins from 16~ms to
16.384~s. Note that the BATSE trigger had one energy band---usually
$\Delta E$=50--300~keV---and the three time bins $\Delta t=$0.064,
0.256, and 1.024~s.  The GBM burst detection algorithms are
described in greater detail in Meegan et al. (2009, submitted).

When the GBM triggers it sends a series of burst alert
packets through the spacecraft and TDRSS to the Earth. Some
of these burst packets, including the burst location
calculated onboard, will also be sent to the LAT to assist
in the LAT's onboard burst detection.  Burst locations are
calculated by comparing the rates in the different
detectors; each the detectors' effective area varies across
the FOV. In addition, the GBM will send a signal over a
dedicated cable to the LAT; this signal will only inform
the LAT that the GBM has triggered.

The continuous GBM data that are routinely telemetered to the ground
can also be searched for bursts that did not trigger the GBM
onboard.  These data will provide rates for all the GBM detectors in
8~energy channels with 0.256~s resolution and in 128~energy channels
with 4.096~s resolution.  In particular, if a burst triggers the LAT
but not the GBM, these rates will at the very least provide upper
limits on the burst flux in the GBM energy band.

%
%
%

\subsection{Onboard LAT Detection}\label{sec:LAT_Onboard}

The LAT flight software will detect bursts, localize them,
and report their positions to the ground through the burst
alert telemetry.  The rapid notification of ground-based
telescopes through GCN will result in multi-wavelength
afterglow observations of GRBs with known high energy
emission.  The onboard burst trigger is described in \cite{Kuehn:07}.

The onboard processing that results in the detection of a
GRB can be subdivided into three steps:  initial event
filtering; event track reconstruction; and finally burst
detection and localization. In the first step all
events---photons and charged particles---that trigger the
LAT hardware are filtered to remove events that are of no
further scientific interest. The events that survive this
first filtering constitute the science data stream that is
downlinked to the ground for further processing. These
events are also fed into the second step of the onboard
burst processing pathway.

The second step of the burst pathway attempts to
reconstruct tracks for all the events in the science data
stream using the `hits' in the tracker's silicon strip
detectors that indicate the passage of a charged particle.
The burst trigger algorithm uses both spatial and temporal
information, and therefore a 3-dimensional track that
points back to a photon's origin is required. Tracks can be
calculated for only about a third the events that are input
to this step, although surprisingly the onboard
track-finding efficiency is 80\% to 90\% of the more
sophisticated ground calculation. However, the onboard
reconstruction is less accurate, resulting in a larger PSF
onboard than on-ground, as is shown by
Fig.~\ref{fig:irf_comparison}. A larger fraction of the
incident photons survive the onboard filtering than survive
the on-ground processing at the expense of a much higher
non-photon background onboard than on-ground; consequently
the onboard effective area is actually larger than the
on-ground effective area, as Fig.~\ref{fig:irf_comparison}
shows.

The rate of events that pass the onboard gamma filter
(currently the same event set that is downlinked and thus
available on-ground) is $\sim$400~Hz.  The rate that events
are sent to the onboard burst trigger, which requires
3-dimensional tracks, is $\sim$120~Hz.  The on-ground
processing creates a transient event class with a rate of
$\sim$2~Hz. Thus onboard the burst trigger must find a
burst signal against a background of $\sim$120 non-burst
events, while on-ground this background is only $\sim$2~Hz.
This difference in non-burst background rate sets
fundamental limits on the onboard and on ground burst
detection sensitivities.

The third step in the burst processing is burst detection,
which considers the events that have passed all the filters
of the first two steps, and thus have arrival times,
energies and origins on the sky. When a detector such as
the GBM provides only event rates, the burst trigger can
only be based on a statistically significant increase in
these rates. However, when a detector such as the LAT
provides both spatial and temporal information for each
event, then an efficient burst trigger will search for
temporal and spatial event clustering. Most searches for
transients bin the events in time and space (if relevant),
but the LAT uses an unbinned method.

The LAT burst trigger searches for statistically
significant clusters in time and space.  The trigger has
two tiers.  The first tier identifies potentially
interesting event clusters for further investigation by the
second tier; the threshold for the first tier allows many
false tier~1 triggers that are then rejected by the second
tier. The first tier operates continuously, except while
the second tier code is running. A GBM trigger is
equivalent to a first tier trigger in that the GBM's
trigger time and position are passed directly to the second
tier.

Tier 1 operates on sets of $N$ events that survived the
first two steps, where currently $N$ is in the range of
40--200.  The effective time window that is searched is $N$
divided by the event rate; for an event rate of 120~Hz and
these values of $N$, the time window is 1/3--5/3~s. Each of
these $N$ events is considered as the seed for a cluster
consisting of all events that are within $\theta_0$ of the
seed; currently $\theta_0=17^\circ$, approximately the 68\%
containment radius of the onboard 3D tracks at low event
energies.  A clustering statistic, described below, is then
calculated for each cluster.  A tier 1 trigger results when
a clustering statistic for any cluster exceeds a threshold
value.  A candidate burst location is then calculated from
the events of the cluster that resulted in the tier 1
trigger.

The onboard burst localization algorithm uses a weighted
average of the positions of the cluster's events. The
weighting is the inverse of the angular distance of an
event from the burst position. Since the purpose of the
algorithm is to find the burst position, the averaging must
be iterated, with the weighting used in one step calculated
from the position from the previous step.  The initial
location is the unweighted average of the events positions.
The convergence criterion is a change of 1~arcmin between
iterations (with a maximum of 10 iterations).  The position
uncertainty depends on the number and energies of events,
but the goal is an uncertainty less that 1$^\circ$. Using
Monte Carlo simulations, this methodology was found to be
superior to others that were tried.

The tier 1 trigger time and localization (or if the GBM
triggered, its trigger time and burst position) are then
passed to the second tier.  Because the second tier is run
relatively infrequently, it can consider a much larger set
of events than the first tier.  Currently 500 events are
considered, which corresponds to a time window of
$\sim$4.2~s. A cluster is then formed from all events in
this set that are within $\theta_2$ ($\sim 10^\circ$) of
the tier 1 burst location. A clustering statistic is then
calculated for this cluster, and if its value exceeds a
threshold, a tier~2 trigger results and the cluster events
are run through the localization algorithm. The resulting
trigger time, burst location and number of counts in four
energy bands are then sent to the ground through the burst
alert telemetry. The second tier is run repeatedly after a
tier 1 trigger in case the burst brightens resulting in a
larger cluster centered on the tier 1 position, and
consequently a tier 2 trigger (if one has not yet occurred)
and a better burst localization (if a tier 2 trigger does
occur).

The clustering statistic is based on the probabilities that
the cluster's events have the observed distances from the
cluster seed position and the arrival time separations,
under the null hypothesis that a burst is not occurring.
Assuming events are thrown uniformly onto a sphere (the
null hypothesis), the probability $p_s$ of finding an event
within $\theta$ degrees of the cluster seed position is
\begin{equation}
p_s = \frac{1-\mathrm{cos}(\theta)}{1-\mathrm{cos}(\theta_m)}
\label{eqn:eventSpatialProb}
\end{equation}
where it is assumed that there are no events at more than
$\theta_m =115^\circ$ (the performance is not sensitive to
this parameter).  Thus for a cluster of $M$ events the
spatial contribution to the clustering statistic is
\begin{equation}\label{eqn:spatialProb}
P_{S} =  \sum_{i=1}^{M}\, |\mathrm{log}_{10}(p_{s_i})| =
   \sum_{i=1}^{M} \Bigg{|} \mathrm{log}_{10}\bigg{(}\,\frac{1-\mathrm{cos}
   (\theta_{i})}{1-\mathrm{cos}(\theta_m)}\bigg{)}\Bigg{|}\; .
\end{equation}

The temporal part of the cluster probability assumes that
the event arrival time follows a Poisson distribution
(again the null hypothesis).  The probability that the
arrival times of two subsequent events differ by $\Delta T$
is
\begin{equation}\label{eqn:eventTemporalProb}
p_t = 1-\exp[-r_t \Delta T] \quad ,
\end{equation}
where $r_t$ is the rate at which events occur within the
area of the cluster. The temporal contribution of each
cluster to the clustering statistic is
\begin{equation}\label{eqn:temporalProb}
P_T = \sum_{i=1}^{M} | \mathrm{log}_{10}(p_{t_i})| = \sum_{i=1}^{M}
  \bigg{|} \mathrm{log}_{10}(1-e^{-r_t \Delta T_i})\, \bigg{|} \;.
\end{equation}

The trigger criterion is
\begin{equation}
\xi  P_T  + P_S > \Theta
\label{eqn:combined_log_prob}
\end{equation}
where $\xi$ is an adjustable parameter that assigns
relative weights to the spatial and temporal clustering,
and $\Theta$ is the threshold.  The two tiers may use
different values of both $\xi$ and $\Theta$.  The overall
false trigger rate depends on the tier 2 value of $\Theta$.

The parameters used by the onboard burst detection and
localization software are sensitive to the actual event
rates, and will ultimately be set based on flight
experience. Currently the thresholds are set high enough to
preclude any triggers, and diagnostic data is being
downlinked and studied.  The thresholds will eventually be
lowered, keeping the false trigger rate at an acceptable
level.

Based on preliminary calculations using a burst population
based on BATSE, we estimate $\sim$1 bursts every two months
will be detected and localized to $1^\circ$ (see Fig.~\ref{fig:sensitivity_like} and Fig.~\ref{fig:localization_like}).

%
%

\subsection{LAT Ground-Based Blind Search}\label{sec:LAT_Ground}

A burst detection algorithm will be applied on the ground
to all LAT counts after the events are reconstructed and
classified to detect bursts that were not detected by the
onboard algorithm, the GBM, or other missions and
telescopes.  Thus this `blind search' is similar to the
first tier of the onboard burst detection algorithm.  The
ground-based search will be performed after each satellite
downlink; to capture bursts that straddle the downlink
boundaries, some counts from the previous downlink are
buffered and used in searching for bursts in the data from
a given downlink. The ground-based blind search algorithm
is very similar to the onboard algorithm described in the
previous section, but will benefit from the full
ground-based event reconstruction and background rejection
techniques that are applied to produce the LAT counts used
for astrophysical analysis. For these data, the particle
background rates will be lower than the onboard rates by at
least two orders-of-magnitude. Furthermore, the
reconstructed photon directions and energies will be more
accurate than the onboard quantities.
Fig.~\ref{fig:irf_comparison} compares the 68\% containment
angle as a function of the photon energy for the onboard
and on-ground LAT count datasets.

\begin{figure}
  \plotone{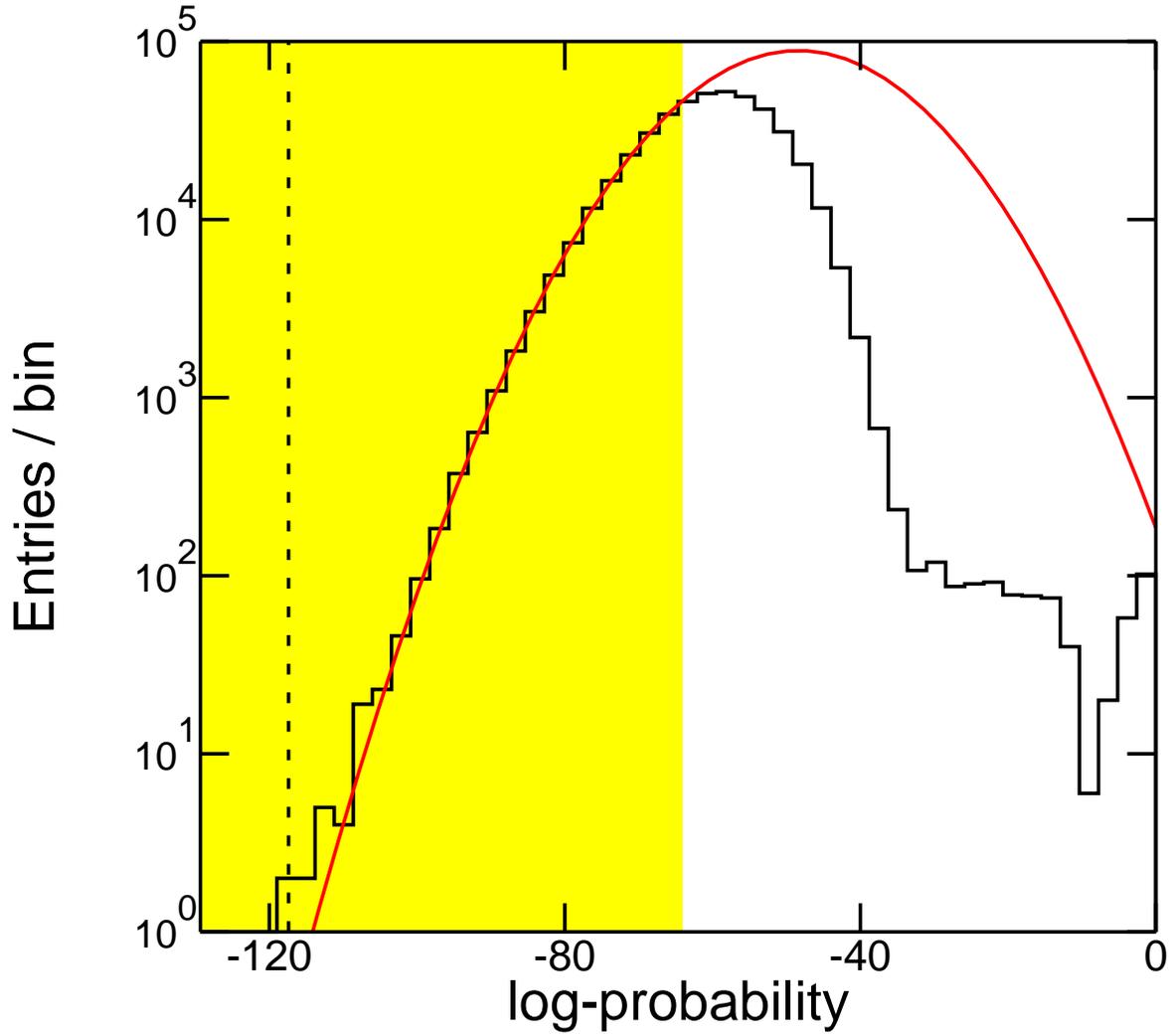}
  \caption{Distribution of log-probability values under the null
  hypothesis obtained from
    applying the ground-based version of the GRB search algorithm to
    sets of 20 counts. The
    shaded region indicates the range over which a Gaussian function,
    shown in red, was fit to these data. The resulting
    5$\sigma$
    threshold at an overall log-probability value of
    $-117$ is plotted as the vertical
    dashed line. Burst candidates are required to have log-probabilities below this threshold.}
  \label{fig:ground_based_null_dist}
\end{figure}

In addition to differing in the reconstruction and
background filtering, the ground-based analysis treats the
input data slightly differently.  The first stage of the
ground-based algorithm is applied to consecutive sets of 20
to 100 counts. As with the onboard algorithm, the number of
counts analyzed is configurable and will be adjusted with
the growth of our knowledge of GRB prompt emission in the
LAT band and of the residual instrumental background.
However, in contrast to the onboard algorithm, the data
sets do not overlap. This ensures that each segment is
statistically independent and generally better separates
the log-probability distributions of the null case (i.e.,
where there is no burst) from the distributions computed
when burst photons are present.
Fig.~\ref{fig:ground_based_null_dist} shows the reference
distribution for the null case derived from simulated
background data. We modeled the low end (large negative
values) of the distribution with a Gaussian, and set the
burst detection threshold at 5$\sigma$ from the fitted
peak. Since this distribution is derived from pre-launch
Monte Carlo simulations with assumed incident particle
distributions and other expected on-orbit conditions, the
thresholds are being re-calibrated with real flight data.
Since we perform an empirical threshold calibration, we can
neglect the constant normalization factors in the
denominators of the single event probabilities shown in
eqs.~\ref{eqn:eventSpatialProb} and
~\ref{eqn:eventTemporalProb}.

\begin{figure}
  \plotone{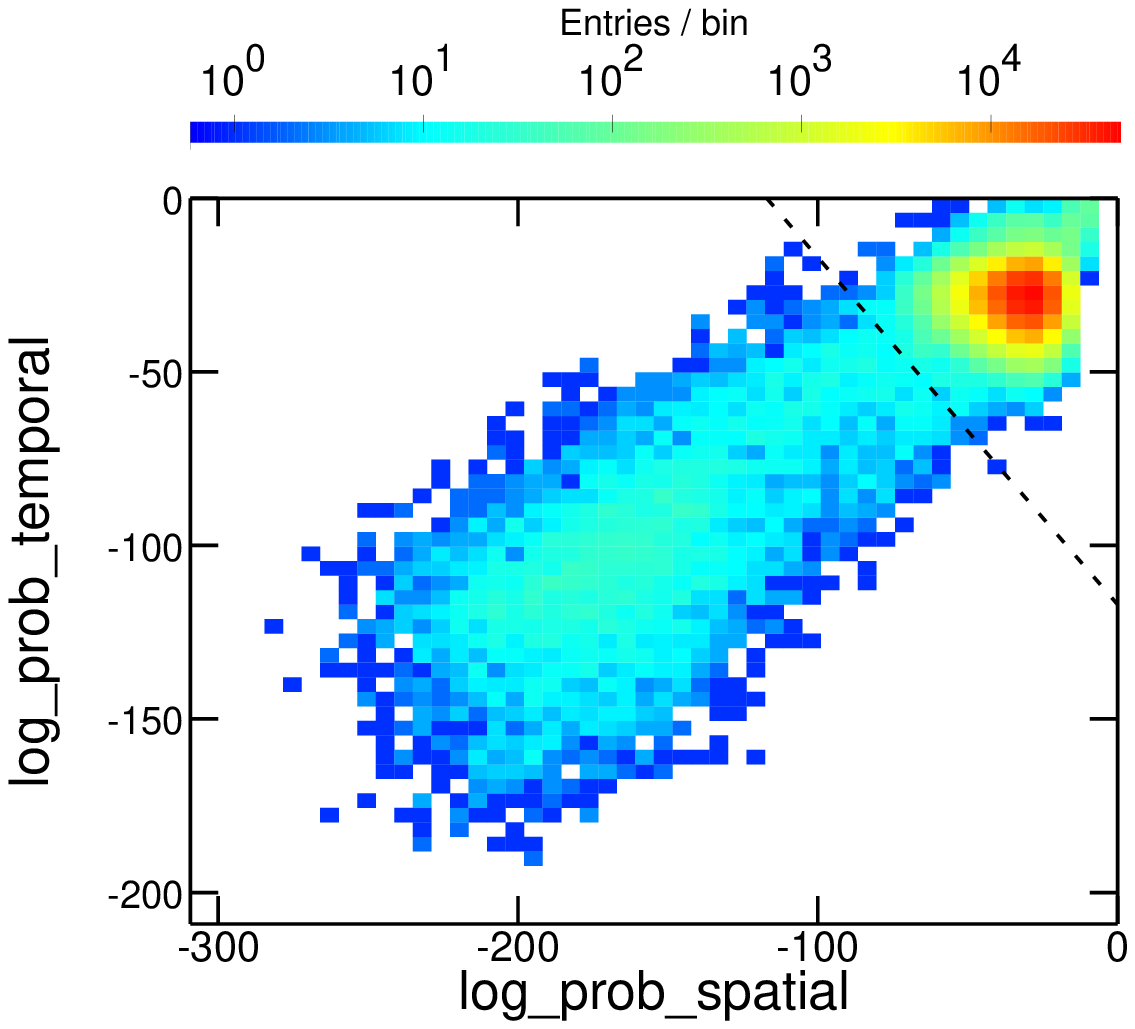}
  \caption{2D histogram of the spatial and temporal log-probability
  components.  The dashed line indicates the
    5$-\sigma$ threshold (an overall log-probability value of
    $-117$) derived from the null distribution
    (figure~\ref{fig:ground_based_null_dist}). Burst candidates are required to lie below this line.}
  \label{fig:logprob_distributions}
\end{figure}

\begin{figure}
  \plotone{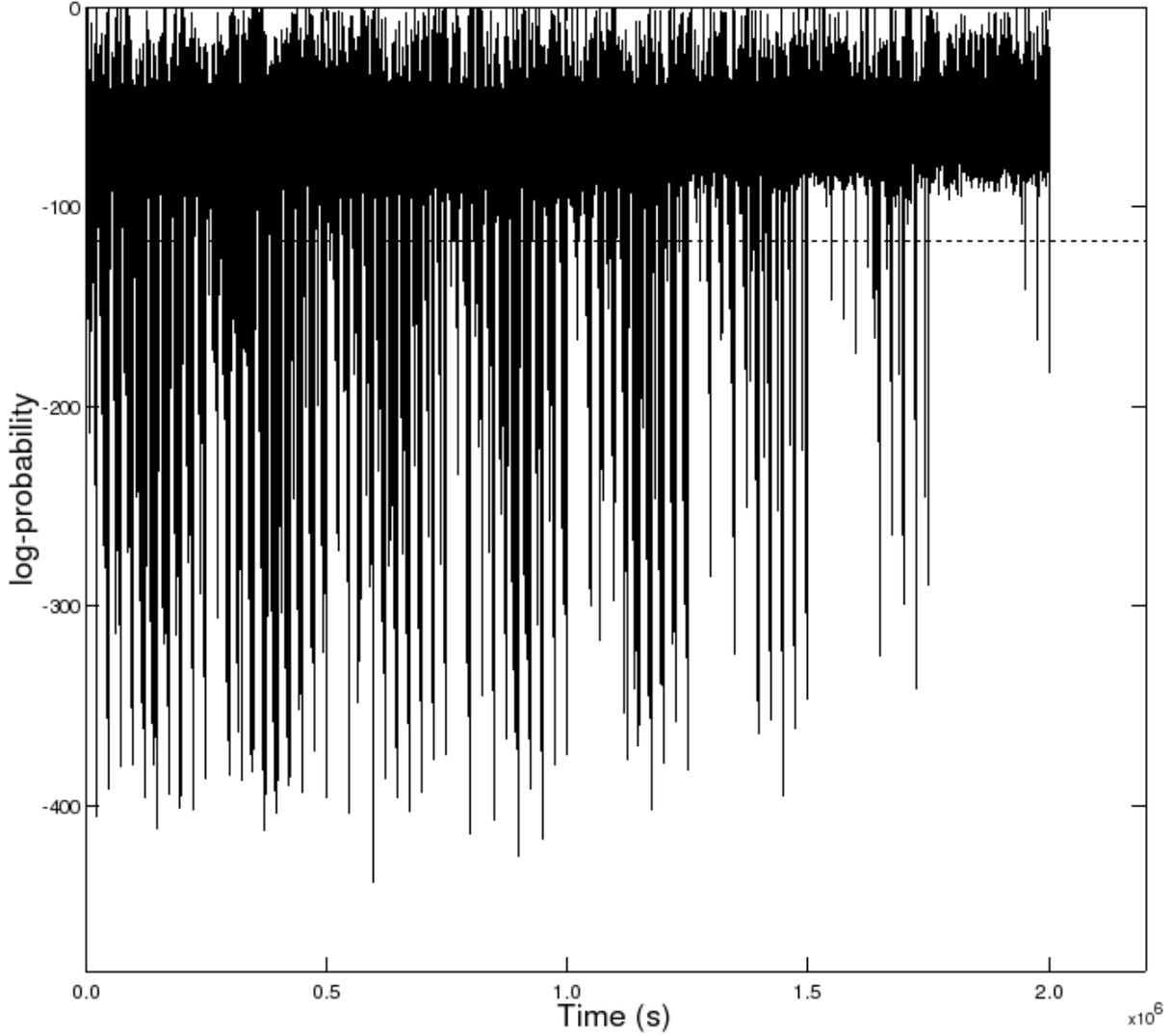}
  \caption{Time history of the ground-based log-probability.  The
  horizontal dashed line shows the 5$\sigma$ threshold derived
  from the Gaussian function fit to the
  log-probabilities distribution under the null hypothesis
  (Fig.~\ref{fig:ground_based_null_dist}). Burst candidates are required to lie below this line.}
  \label{fig:logprob_time_history}
\end{figure}

The overall log-probability is the sum of spatial and
temporal components (see eq.~\ref{eqn:combined_log_prob}),
which we weight equally ($\xi$=1).
Fig.~\ref{fig:logprob_distributions} shows the 2D
distributions for the temporal and spatial components.  The
dashed line in Fig.~\ref{fig:logprob_distributions}
corresponds to the 5$\sigma$ threshold with this weighting.
Fig.~\ref{fig:logprob_time_history} shows the time history
of the log-probabilities as applied to the GRB grid data.
The excursions across the threshold line indicate the burst
candidates.

While the onboard burst trigger performs two passes through
the data with the temporal-spatial clustering likelihood
algorithm, the ground-based detection analysis performs
only one such pass. If a candidate burst is found in the
ground-based analysis, counts from a time range bracketing
the trigger time undergoes further processing to determine
the significance of the burst.  If the burst is
sufficiently significant, it is localized and its spectrum
is analyzed.  These analyses use the unbinned maximum
likelihood method that is applied to LAT point sources.

\subsection{GRB Candidate Follow-up Processing}

When a candidate burst location and trigger time is
provided by the ground-based blind search, a LAT or GBM
onboard trigger, or another burst detector such as {\it
Swift}---we will call this a first stage detection---a LAT
ISOC data processing pipeline will analyze the LAT counts
to determine the significance of a possible LAT detection.
This step in deciding whether the LAT has detected a burst
is similar to the tier two analysis of the onboard
algorithm. If the LAT has detected a burst, the pipeline
will localize the burst and determine its temporal start
and stop.  All of the analyses described in this section
will be performed using the ``transient'' class. These data
selections have a larger effective area at a cost of
somewhat higher instrumental background, particularly in
the 50--200 MeV range.  For bright transients, such as are
expected for GRBs, this trade-off is advantageous given the
short time scales.

\begin{figure}
  \plotone{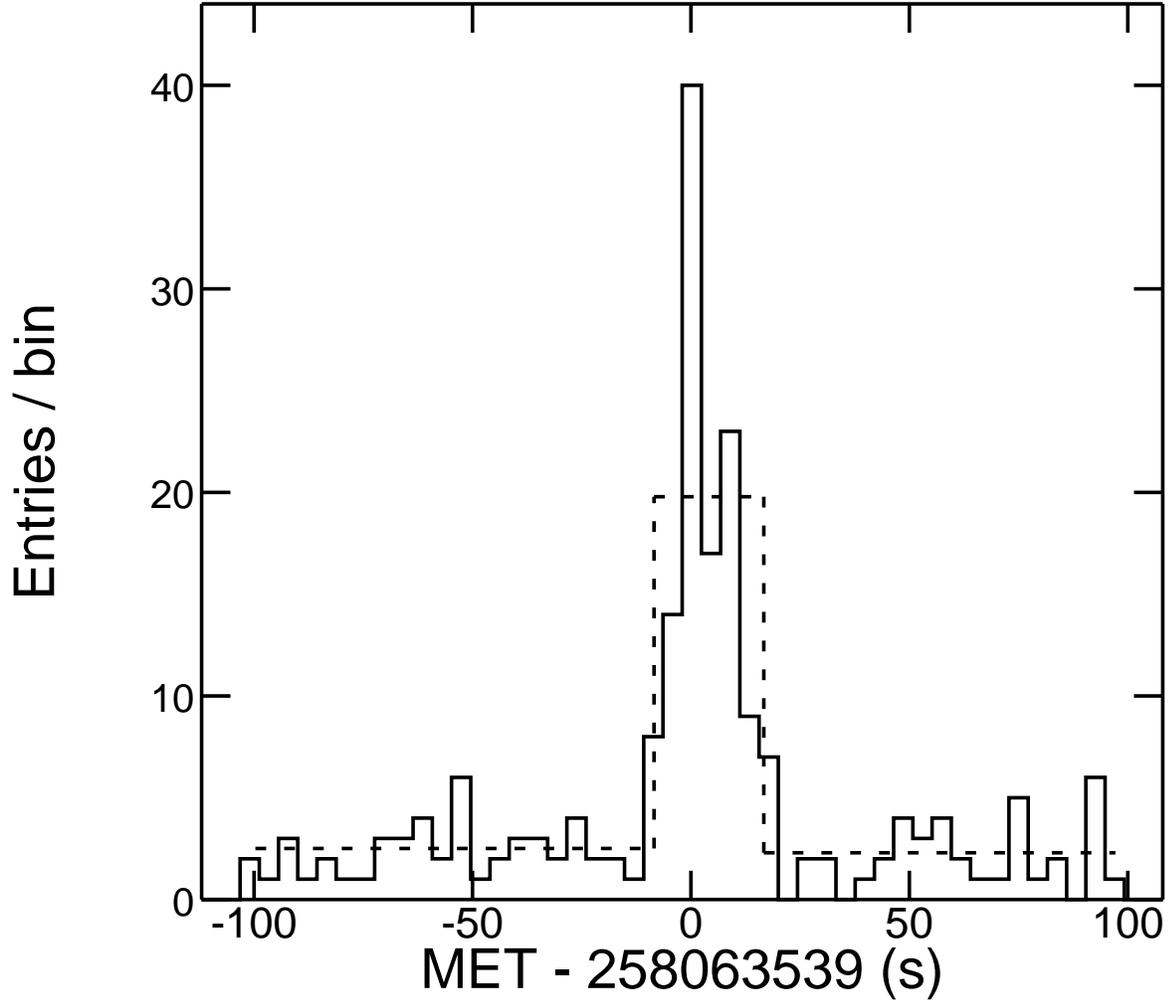}
  \caption{LAT counts light curve for a simulated burst (solid
    histogram) and a piece-wise constant light curve derived using
    the Bayesian Blocks analysis of the event arrival times (dashed
    histogram).}
  \label{fig:ASP_light_curve}
\end{figure}

The first step in the follow-up processing is determining
the time interval straddling the candidate burst during
which the LAT count rate is greater than the expected
background rate. The counts are selected from a $15^\circ$
acceptance cone centered on the candidate burst position
and from a 200 second time window centered on the candidate
burst trigger time.  This time window is designed to
capture possible precursor emission that may be present in
the LAT band. Both the acceptance cone radius and the time
window size are configurable parameters in the processing
pipeline. With this acceptance cone radius, the total event
rate from non-GRB sources is expected to be $<0.1$ to 0.5
Hz for normal scanning observations, depending on how far
the candidate position is from the brightest parts of the
Galactic plane emission. The event arrival times are
analyzed using a Bayesian Blocks algorithm
\citep{Jackson:03,Scargle:98} that aggregates arrival times
in blocks of constant rate and identifies ``change points''
between blocks with statistically significant changes in
event rate. The burst start and stop time are identified as
the first and last change points from the resulting light
curve. An example of the results of this analysis is shown
in Fig.~\ref{fig:ASP_light_curve}.

If no change points are found within the 200 second
bracketing time window, then the counts from the first
stage time window and burst position will be used in
calculating upper limits.  In these cases, the position
refinement step will be skipped and background model
components will be included in the significance and upper
limits analysis.

\begin{figure}
  \plotone{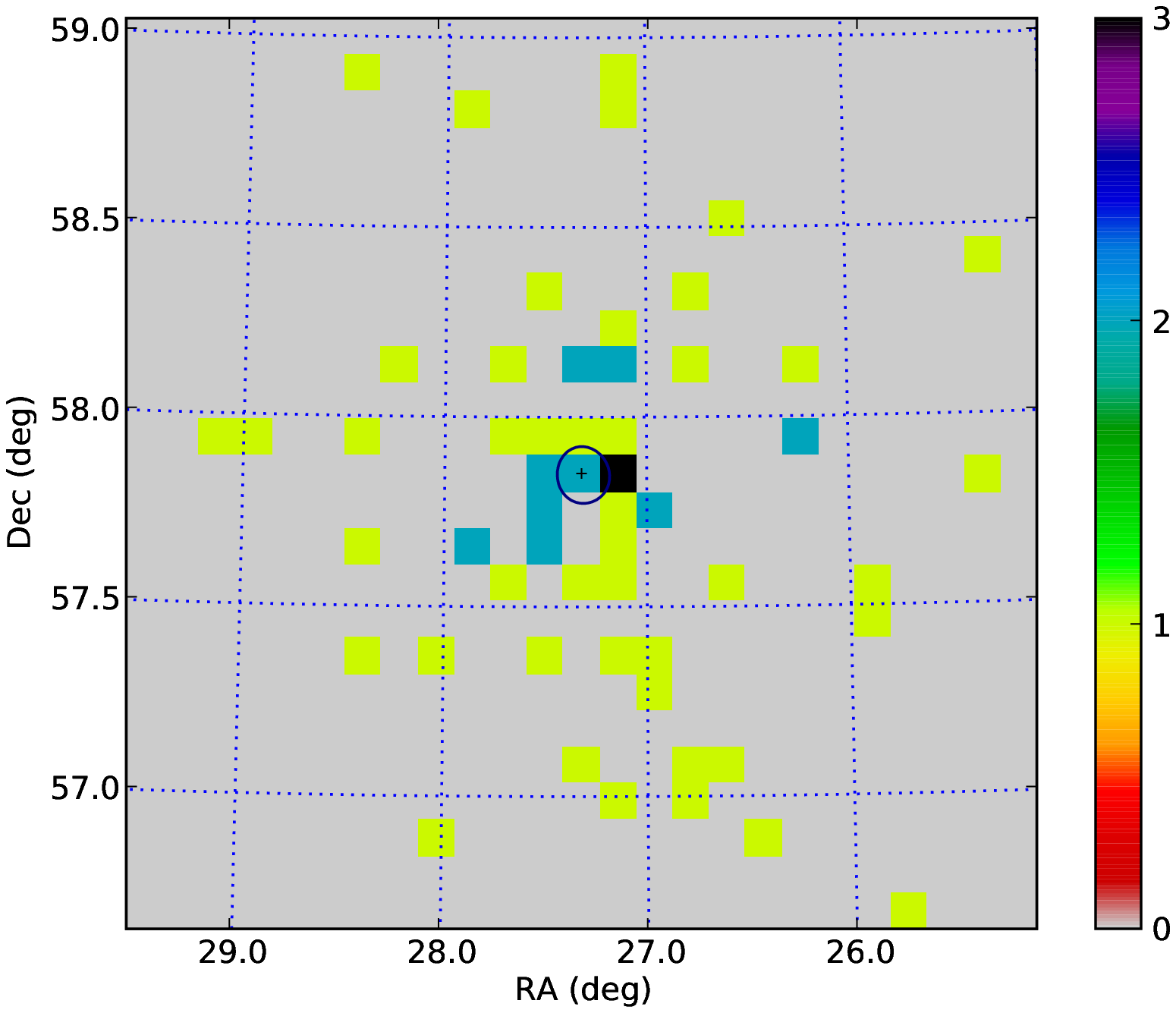}
  \caption{LAT counts map for the simulated burst in
    Fig.~\ref{fig:ASP_light_curve} using only the counts between the first
    and last change points. The best-fit position and 90\%
    error contour derived from the maximum likelihood analysis are
    overlaid. The color scale on the right shows the counts per pixel.}
  \label{fig:ASP_cmap_ts_contours}
\end{figure}

If application of the Bayesian Block algorithm to the LAT
arrival times finds a statistically significant increase in
the count rate above background, i.e., if at least two
change points were found, then further analysis uses only
the counts between the first and last change points to
exclude background.  The position is refined with the
standard LAT maximum likelihood software that folds a
parameterized input source model through the instrument
response functions to obtain a predicted distribution of
observed counts.  The parameters of the source model are
adjusted to maximize the log-likelihood of the data given
the model. For these data, the background counts are
sufficiently small that a model with the different
background components usually used in point source analysis
is not needed, and a model with a single point source
should suffice to localize the burst. The burst spectral
parameters and burst coordinates are adjusted within the
extraction region to maximize the log-likelihood, and the
best-fit position is thereby obtained.  Error contours are
derived by mapping the likelihood surface in position
space, with 90\% confidence limit (CL) uncertainties given
by the contour corresponding to a change in the
log-likelihood of 2.305. This value is equal to
$\Delta\chi^2/2$ for 2 degrees-of-freedom (dof).
Fig.~\ref{fig:ASP_cmap_ts_contours} shows an example counts
map with the 90\% CL contour overlaid.

For spectral analysis and the definitive burst significance
calculation we use the counts within the first and last
change points and at the center of a $15^\circ$ radius
acceptance cone around the maximum likelihood position.
Again we use maximum likelihood to derive the basic burst
parameters from the LAT data alone. Since this is an
automated procedure, a simple power-law model is chosen as
the default. For brighter bursts, background model
components are not needed. For fainter bursts, such as
those burst candidates for which we only have a first stage
detection, including the background is essential to
determine the significance of a faint burst in the LAT data
and for deriving upper limits.

\subsection{Quantifying Significance and Upper
Limits}\label{sec:UL}

As discussed in \S~\ref{sec:Semi_Analytical_Sensitivity},
the likelihood ratio test (LRT) is a natural framework for
hypothesis testing, and we will use this method for
quantifying the significance of a candidate burst.  The
background models used for the null hypothesis (i.e., that
a burst is not present) can be simplified considering the
expected number of counts from each background component
over the short GRB time scales ($< {\cal O}(10^2)$~s). For
determining the significance of a source, we compute the
test statistic defined in eq.~\ref{eq:Ts}.  We are fairly
conservative and require a $T_s > 25$, corresponding to
5$\sigma$ for 1 dof, in order to claim a detection.

Upper limits may be computed in several different ways.  A
method that has been used in the past for GRBs and other
transient astronomical sources is a variant of the
classical ``on source-off source'' measurement. In this
method, one defines an appropriate background interval
prior to the time of the candidate burst, and using the
inferred background levels, one derives an upper-limit for
the source flux given the counts that are observed during
the interval containing the candidate burst.  Application
of this procedure requires that the observing conditions
(instrument response, intrinsic background rates, etc.)
during the background interval be sufficiently similar to
those for the interval containing the putative signal.  For
the short time spans appropriate for GRBs ($\la 100$ s),
simulations have shown that the instrumental background
rates are fairly constant; in survey mode, at fixed rocking
angle, the LAT FOV scans across the sky at a few degrees
per minute, so the instrument response to a given source
location will be roughly constant as well.  A major benefit
of this procedure is that it is model-independent. However,
being model independent, it is also fairly conservative;
and in general, it will not give the most constraining
upper-limit.

A more stringent upper-limit may be computed with the
``profile likelihood'' method.  In this method the
normalization of the source flux (or a parameter that
determines this normalization) is varied while fitting all
the other model parameters, resulting in the variation of
the log-likelihood (the fitting statistic) as a function of
the source normalization.  For a two-sided interval, under
Wilks' theorem the 90\% confidence region corresponds to a
change in the log-likelihood from the extremum of 2.71/2,
i.e., $=\Delta\chi^2/2$ for 1 dof.  For a one-sided
interval, as in the case of an upper-limit, this
corresponds to a 95\% CL.

\begin{figure}
  \plottwo{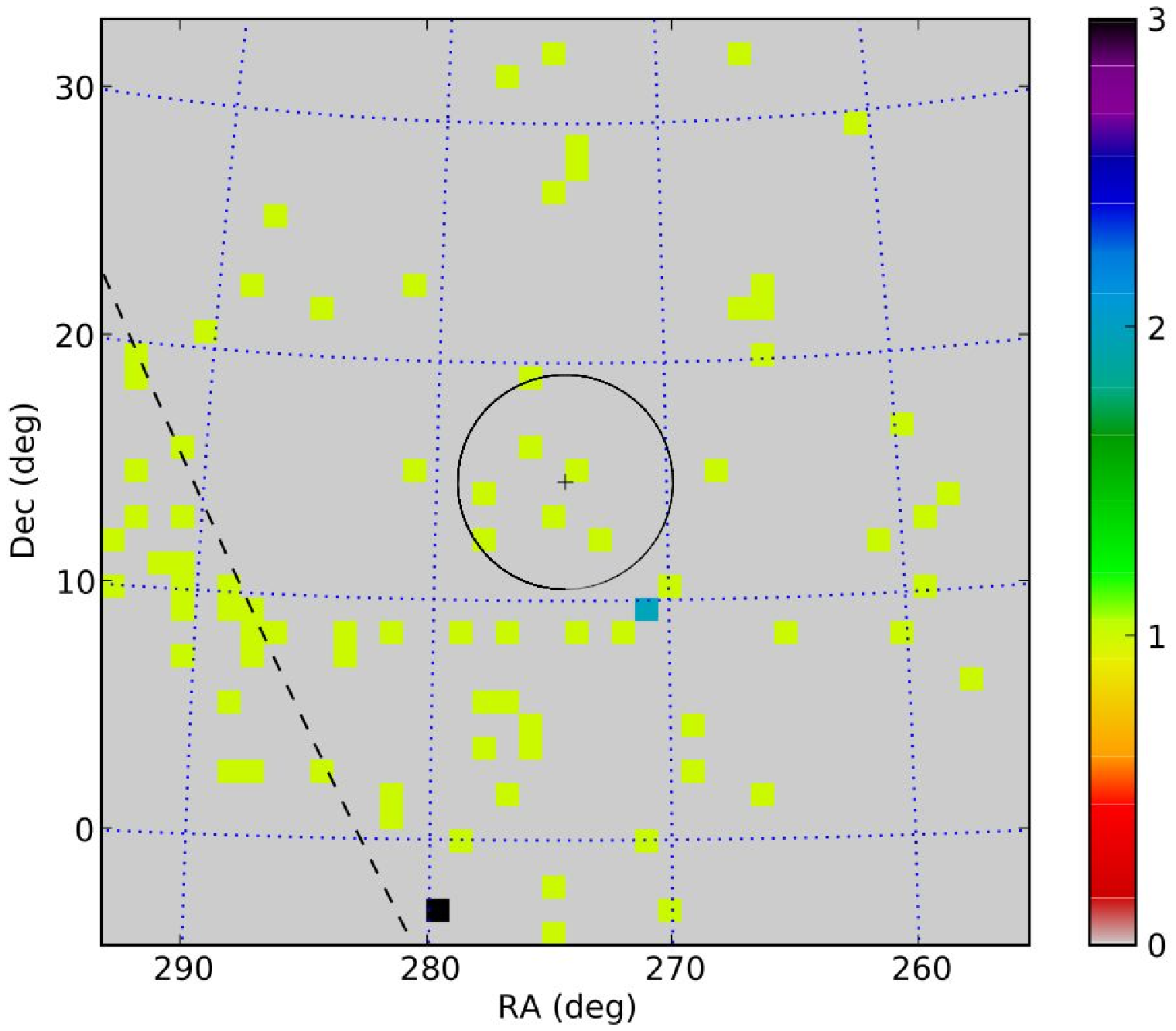}{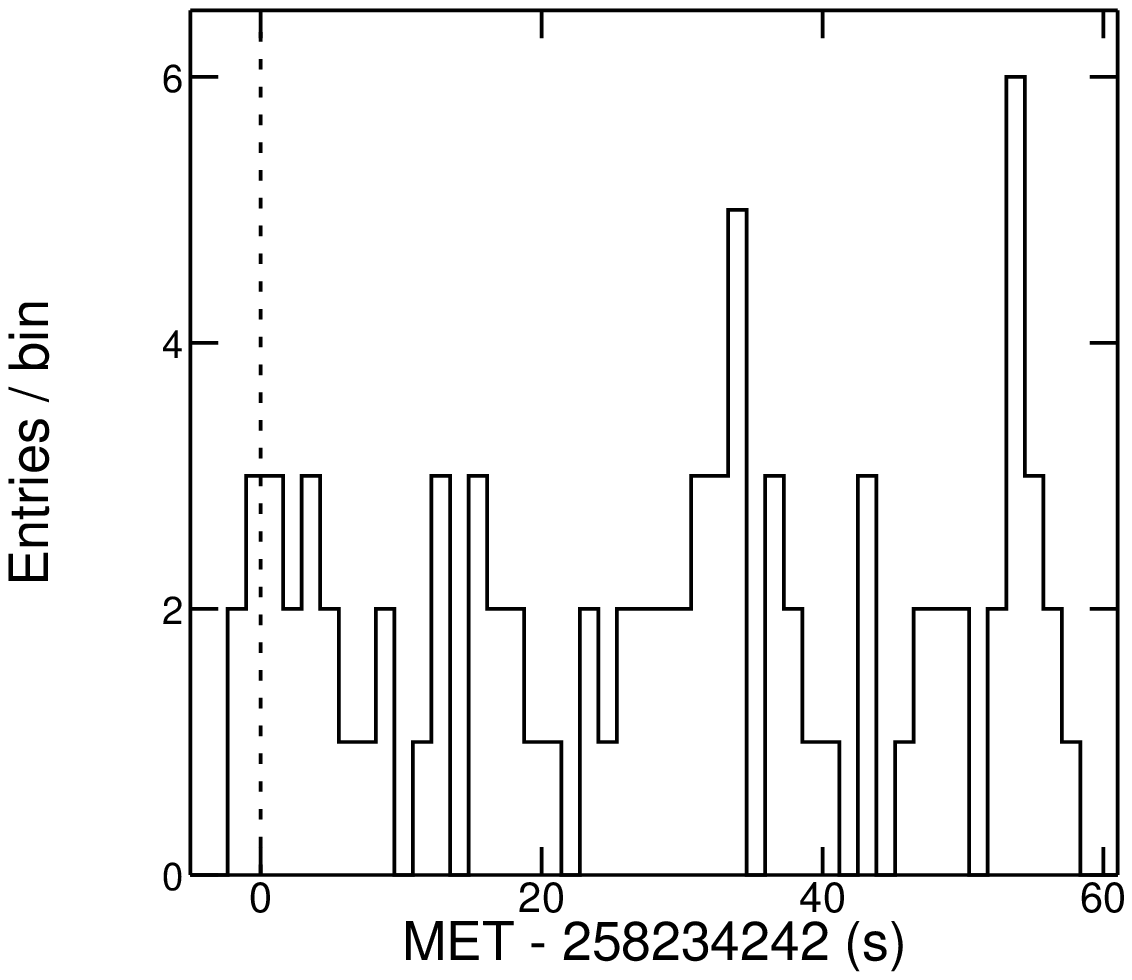}
  \caption{{\bf Left:} LAT counts map for a 60 s time window containing
    the GBM trigger time of a simulated burst.  The GBM location and
    $4.5^\circ$ error circle are plotted.  The dashed line indicates
    the location of the Galactic plane. The color scale on the right shows the counts per pixel. {\bf Right}: Counts light curve for
    these data.  The GBM trigger time is indicated by the vertical
    dashed line.}
  \label{fig:grb_ul_cmap_lc}
\end{figure}

\begin{figure}
  \plotone{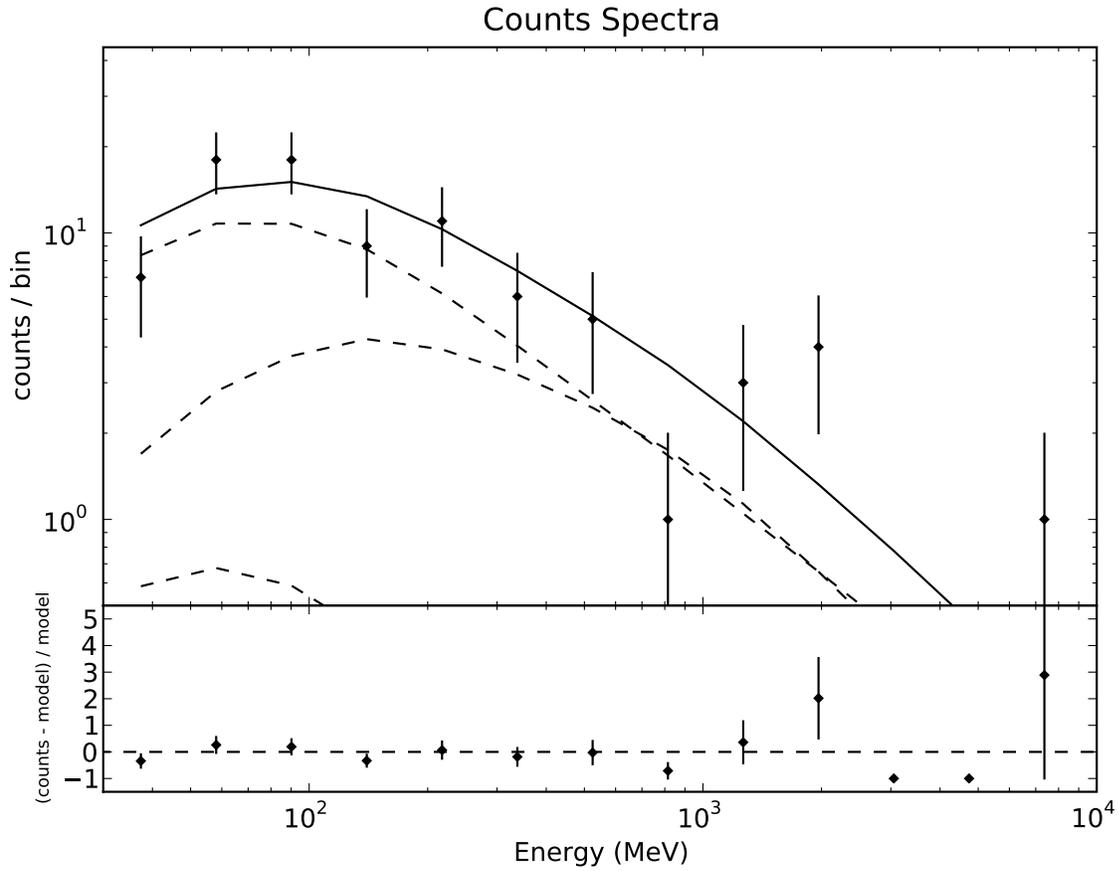}
  \caption{Fitted counts spectrum and residuals for the data shown in
    Fig.~\ref{fig:grb_ul_cmap_lc}.  The contributions of the three
    model components are plotted as the long dashed curves, and from
    top to bottom, are the Galactic diffuse, extragalactic diffuse,
    and point source.  The solid curve is the sum of the three
    components.}
  \label{fig:grb_counts_spectrum}
\end{figure}

\begin{figure}
  \plotone{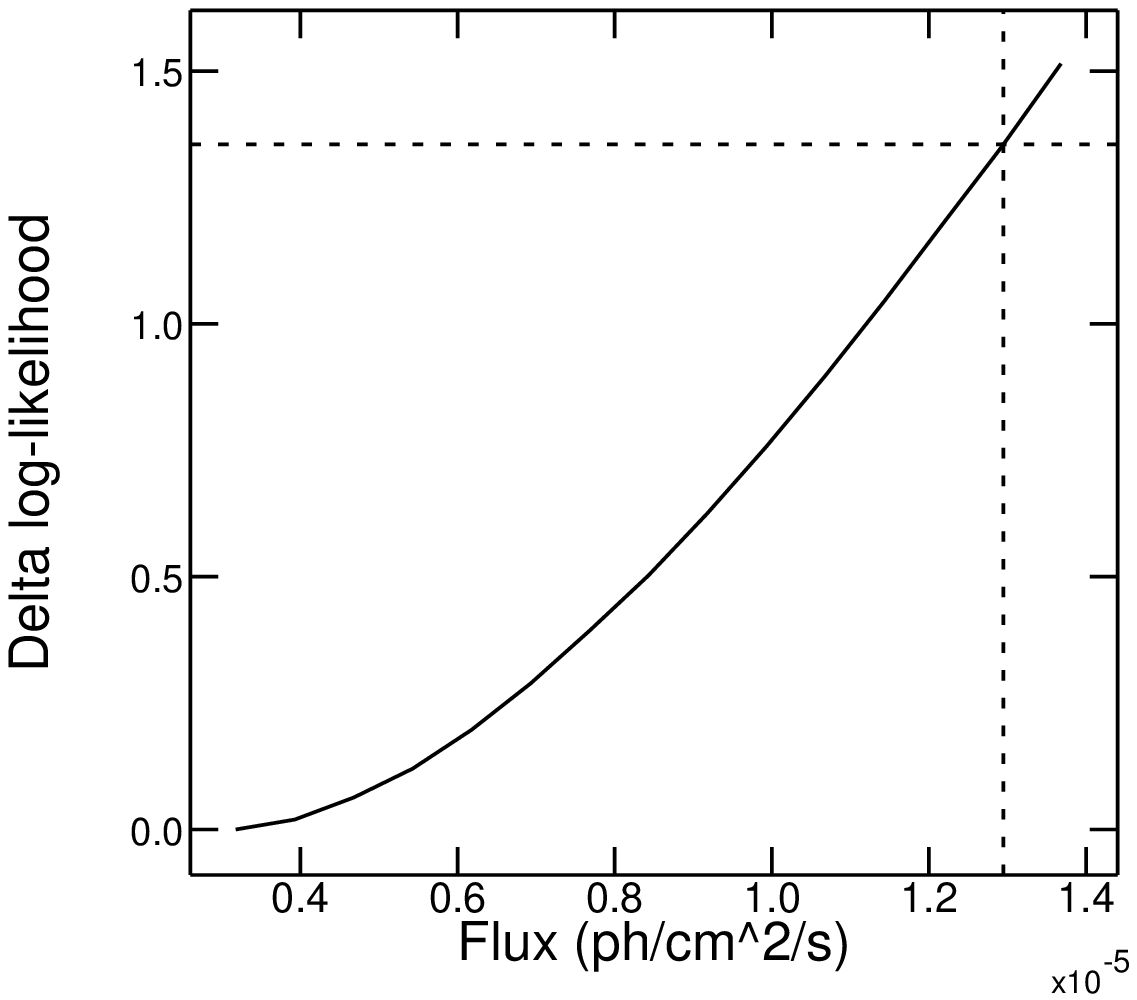}
  \caption{Change in the log-likelihood as a function of GRB flux for
    $E > 100$\,MeV. The horizontal dashed line indicate the 95\% CL
    corresponding to an upper-limit of $1.3\times 10^{-5}$
    ph~cm$^{-2}$~s$^{-1}$.}
  \label{fig:likelihood_profile}
\end{figure}

To illustrate the method, we apply this analysis to
simulated data.  Fig.~\ref{fig:grb_ul_cmap_lc} shows a LAT
counts map and lightcurve for the time and location of a
simulated burst that was detected in the GBM, but is not
evident in the LAT data.  The best-fit flux and error
estimate for a point source is $3.2 \pm 4.5 \times
10^{-6}$\,ph~cm$^{-2}$~s$^{-1}$ for energies $E
> 100$\,MeV. The test statistic for the point source is
$T_s = 0.67$, consistent with the flux measurement's large
error bars and the lack of a burst detection.
Fig.~\ref{fig:grb_counts_spectrum} shows the fitted counts
spectrum and residuals from this fit.
Fig.~\ref{fig:likelihood_profile} shows the change in
log-likelihood as a function of scanned flux value. For a
95\% CL upper limit, we find a value of $1.3 \times
10^{-5}$\,ph~cm$^{-2}$~s$^{-1}$.

\begin{figure}
  \plottwo{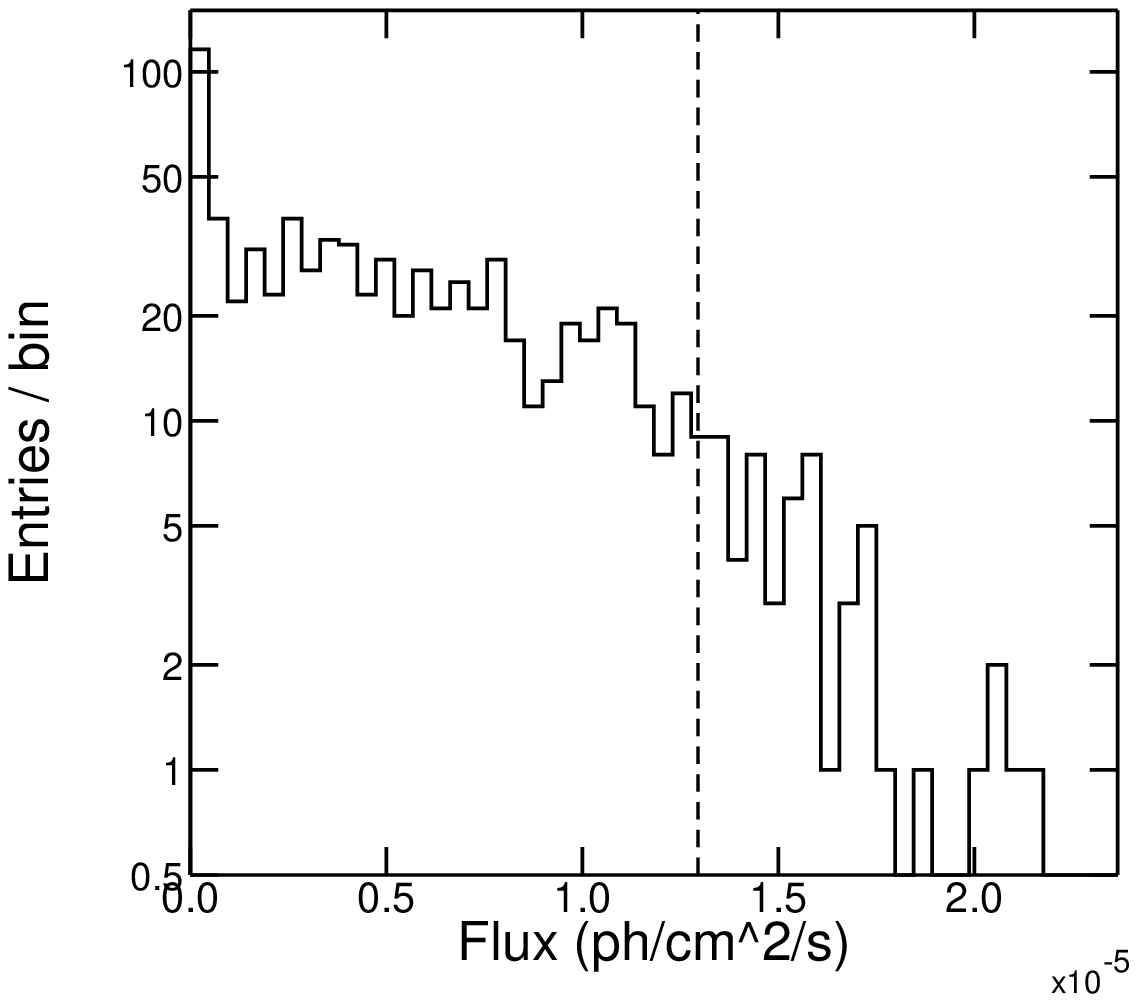}{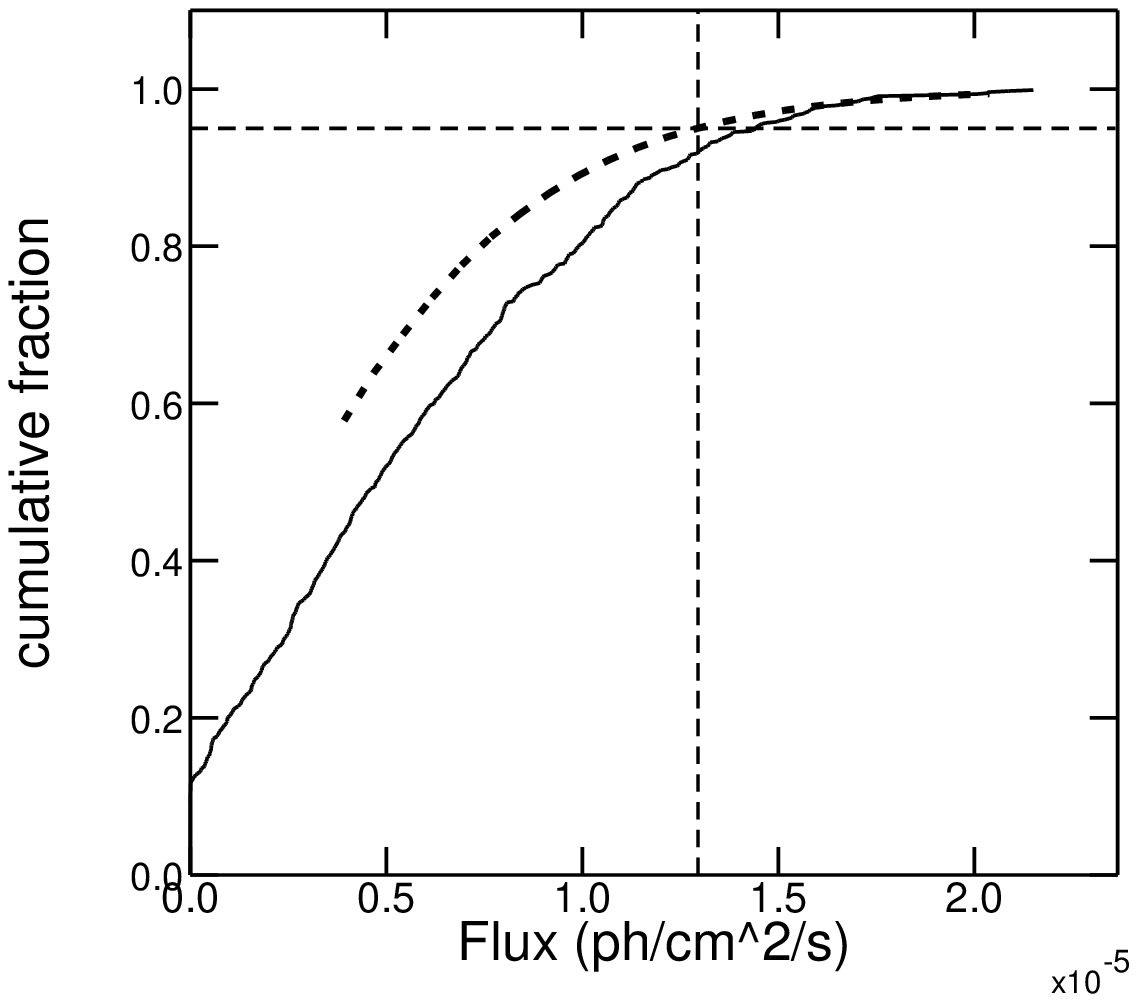}
  \caption{{\bf Left}: Distribution of fitted fluxes for the point
  source representing the GRB derived from 766 LAT simulations using
  the best-fit model obtained from the original dataset. {\bf Right}:
  The solid curve is the normalized cumulative distribution determined
  from the fitted flux distribution.  The dotted curve is the cumulative
  fraction that would be predicted by the likelihood profile shown in
  figure~\ref{fig:likelihood_profile}.}
  \label{fig:grb_mc_results}
\end{figure}

To check the method's validity, we ran Monte Carlo
simulations under the same observing conditions and using
the source model and best-fit parameters from the
likelihood analysis as inputs, and we analyzed each
simulation to find the best-fit flux. The left panel of
Fig.~\ref{fig:grb_mc_results} shows the distribution of
fitted fluxes for these simulations, and the right panel
shows the normalized cumulative distribution for these data
and the cumulative distribution inferred by computing the
corresponding $\chi^2$ probability from the profile
likelihood curve shown in
Fig.~\ref{fig:likelihood_profile}.

%
%
%
%

%
%

\section{Spectral Analysis}\label{sec:Spectral_Analysis}

To demonstrate the spectral analysis that will be possible with the
$Fermi$ data, we present two sample analyses, the first the joint fit
of GBM and LAT count spectra, and the second the search for a cutoff
in the LAT energy band. In both cases we use transient class LAT
counts.  In general, bursts are short but bright, and thus we can
tolerate the higher background rate of the transient class to
increase the number of burst counts.  While we focus here on LAT-GBM
joint fits, such fits will also be possible between the $Fermi$
detectors and those of other missions, such as {\it Swift}
\citep{Stamatikos:08a,Band:08}.

\subsection{GBM and LAT Combined Analysis}

\begin{figure}
  \plottwo{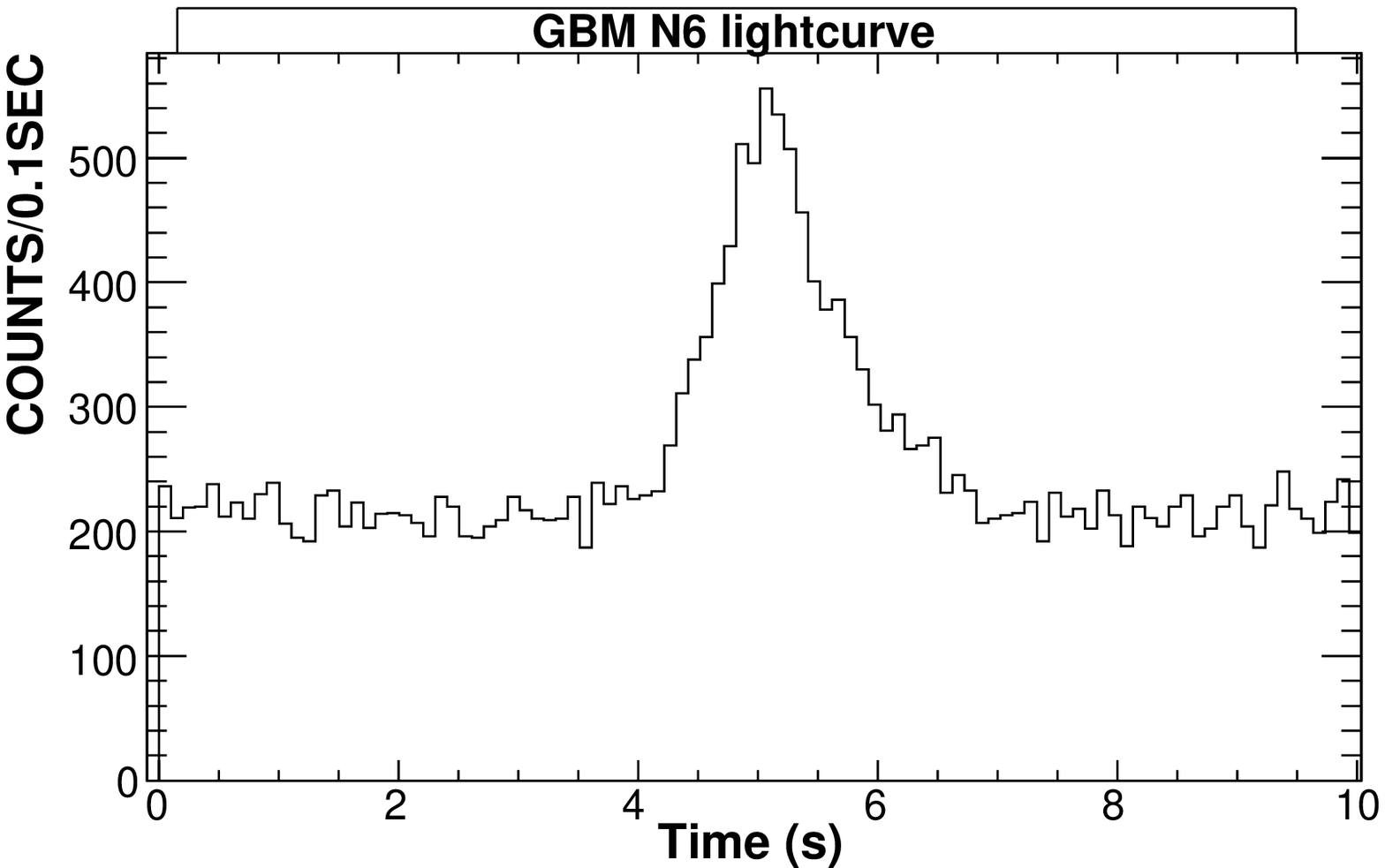}
  {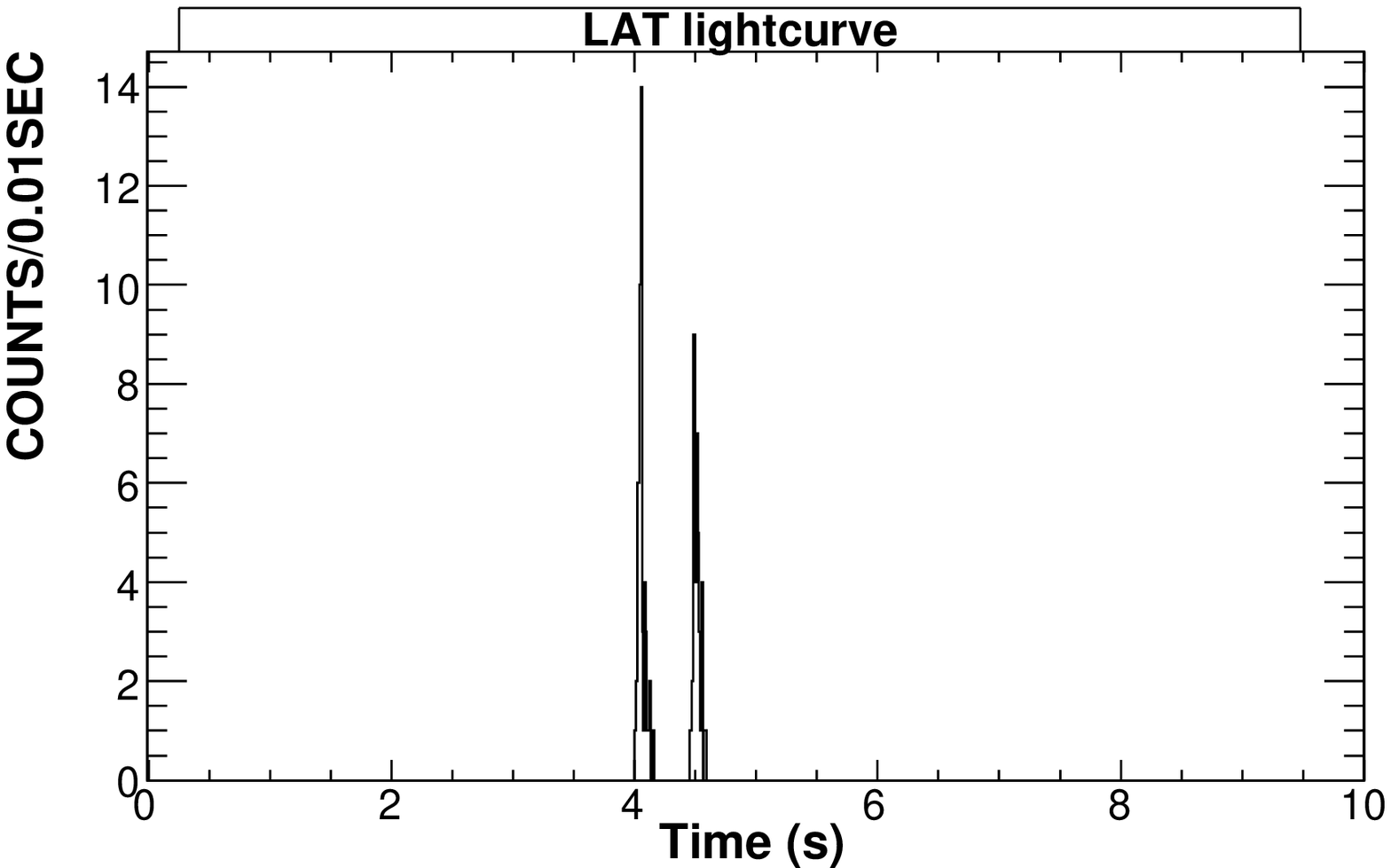}
  \caption{GBM NaI (left) and LAT (right) light curves of the prompt
  emission from the simulated burst.}
  \label{fig:GRB090128A-1}
\end{figure}

In this example, we assume that a simulated burst was detected and 
localized by the GBM.  Analysis of
the LAT data found 160 transient event class photons in a
20$^\circ$ region surrounding the GBM position during the
3~s prompt phase observed by the GBM; the Automated Science
Processing (ASP) that will be run after the LAT events are
reconstructed (\S \ref{sec:LAT_Description}) localized the
burst with an uncertainty radius of $0.05^\circ$.
Fig.~\ref{fig:GRB090128A-1} shows the GBM and LAT light
curves.

\begin{figure}
  \plotone{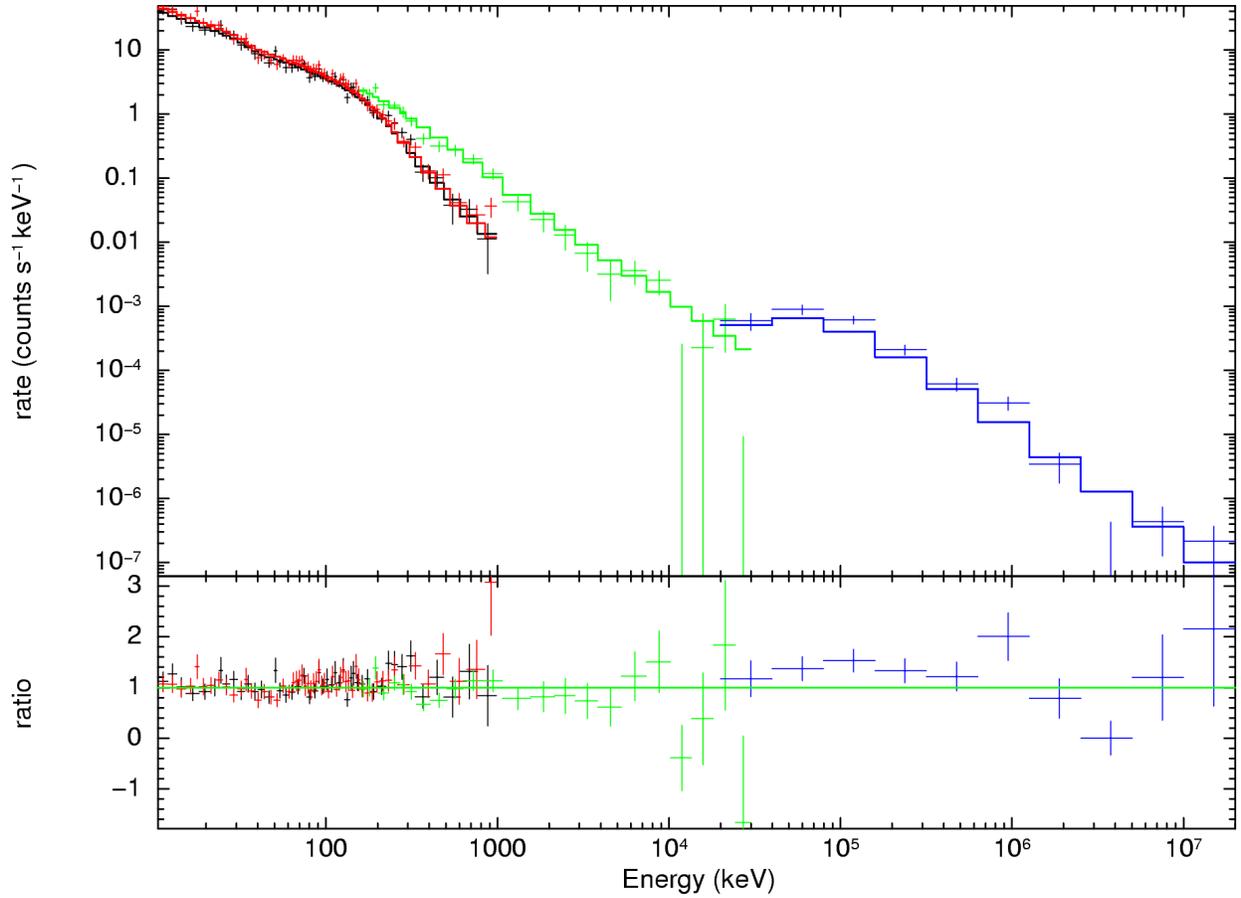}
  \caption{Photon spectrum of the simulated burst: in the top panel, crosses show the data of the different sub-detectors (two NaI detectors in black and red, one BGO in green, and the LAT in blue) and the histogram denotes the best fit of a Band function. The bottom panel shows the ratio of the simulated data to the fit model.}
  \label{fig:GRB090128A-2B}
\end{figure}

The simulated GBM and LAT data, both event lists, were
accumulated over the burst's prompt phase, and the LAT
events were binned into 10 energy bins.  Two NaI and one
BGO detector provided count spectra. The GBM background
spectra used to simulate the counts were used as the
background for the GBM count spectra, while the LAT data
were assumed not to be contaminated by background events.
We performed a joint fit to the 4 count spectra (from 2
NaI, one BGO and the LAT detectors) with the standard X-ray
analysis tool XPSEC using the Cash statistic
\citep{Cash:79}. The `Band' spectrum (eq.~\ref{eqn:BandF2})
was used to create the simulated data and for the joint
fit. Fig.~\ref{fig:GRB090128A-2B} shows the simulated data
(with error bars) and best-fit model (histogram).  The fit
yielded $\alpha = -0.97 \pm 0.05$ (input value of $-1.09$)
and $\beta = -1.80 \pm 0.01$ (input value of $-1.90$).

Thus $Fermi$ will measure the energy spectrum of bursts over 7 orders
of magnitude in energy through its combination of detectors. The
energy bands of the NaI and BGO detectors overlap in the energy
region of the peak energy, and the BGO and the LAT energy bands also
overlap.

\subsection{Study of GRB high-energy properties with the LAT}

Whether the burst spectrum is a simple power law in the LAT
energy band, or has a cutoff spectrum is of great
theoretical interest (see \S~\ref{sec:HE_abs}). Therefore,
we simulated and then fit spectra with such cutoffs to
determine if they would be detectable.

We used the simulation software described in
\S~\ref{sec:Phenomenological} to simulate {\nicola 5 years} of $Fermi$
observations. In this simulation, the temporal and spectral
properties of GRBs were based on a phenomenological or physical
model, including not only synchrotron emission but also inverse
Compton emission for a few bursts. The simulated spectra did not
have any intrinsic cutoffs, but included gamma-ray absorption by the
Extragalactic Background Light (EBL) between the burst and the
Earth, following the model of \citet{Kneiske:04}. This extrinsic
cut-off only appears at the highest energies (at least 10~GeV),
depending on the distance of the bursts.

The search for high-energy cut-offs was performed using
only simulated LAT data. First we selected those bursts
that have no inverse Compton component, and more than 20
LAT counts. Each count spectrum was fit both by a simple
power law and by a power law with an exponential cutoff
with characteristic energy $E_c$.

\begin{figure}
  \plotone{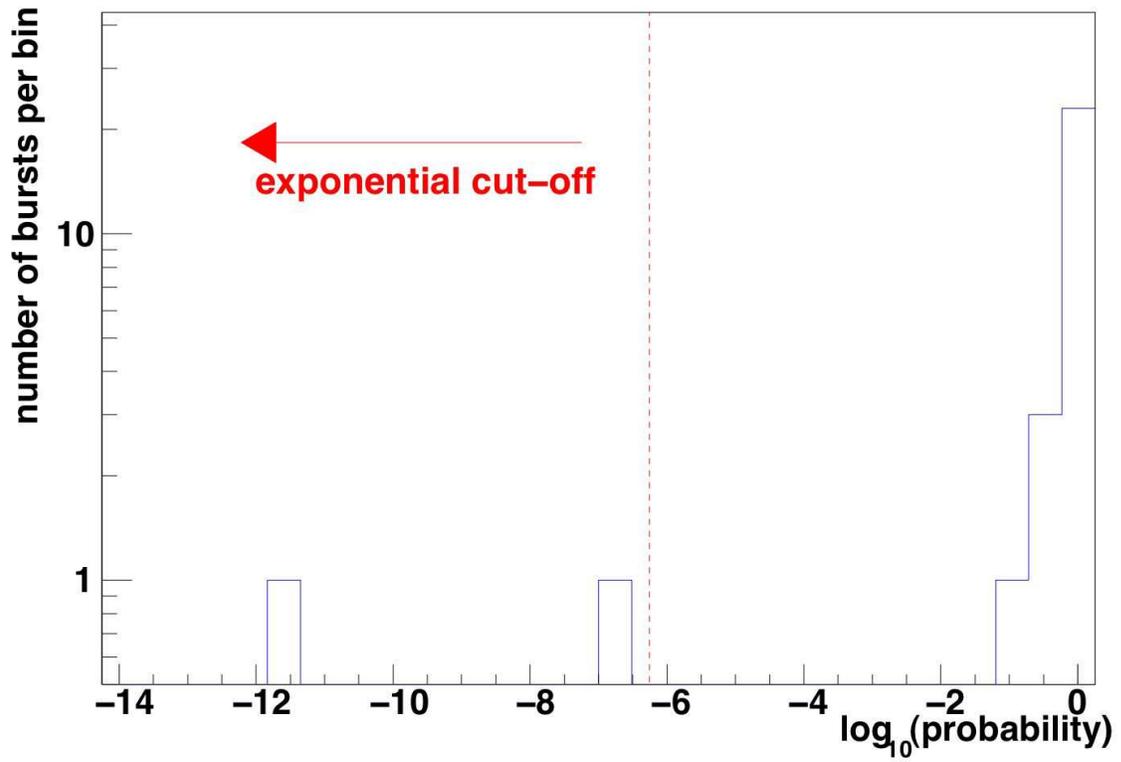}
   \caption{
$\chi^2$-probability of the difference of the likelihoods of
  fits of a power law with and without an exponential cutoff: a
  probability of $<5.7 \times 10^{-7}$ corresponds to a 5$\sigma$
  detection of a cutoff.
}
  \label{fig:cutoffA}
\end{figure}

The likelihoods of the resulting fits were examined to
evaluate the improvement of the fit by adding the cutoff
(one additional parameter). The difference of the
likelihoods follows a $\chi^2$-distribution with one degree
of freedom, with the null hypothesis probability
distribution shown in Fig.~\ref{fig:cutoffA}. Two bursts
exhibit a very small probability of being consistent with
no cutoff, and thus we consider these bursts to have a
statistically significant high-energy cutoff. While both
bursts have average redshifts (1.71 and 3.35) compared to
the full sample, they are very bright, with more than 1000
photons detected.  

For these two bursts we performed a second fit using the
parameterisation of the EBL cut-off proposed
by~\citet{Reyes:07} where the cutoff is $\exp(-\tau)$, with
$\tau=1+(E-E_1)/P$ for $E>E_1-P$, and $0$ otherwise; $E_1$
is the redshift-dependent energy where the optical depth is
unity, and $P$ is a redshift-dependent energy scaling
factor.
The two fitted values of $E_1$ ($51.5^{+6.7}_{-3.6}$ GeV and $43.5^{+31.0}_{-10.0}$ GeV),
are in good agreement with the true values ($46.6$ GeV and $30.7$ GeV) of the model used for the simulation.
Thus the LAT will be sensitive to cutoffs in the brightest bursts, with good spectral reconstruction.

%
%

\section{Coordination with Other Burst Missions}\label{sec:Other_Missions}

\subsection{Coordination with {\it Swift}}\label{sec:Swift}

The $Fermi$ detectors will provide few localizations accurate to less
than 10~arcmin that are necessary for the optical followups that can
determine redshifts.  On the other hand, the {\it Swift} instruments
\citep{Gehrels:04}---the Burst Alert Telescope (BAT), the X-Ray
Telescope (XRT) and the Ultraviolet-Optical Telescope
(UVOT)---provide progressively better burst localizations that are
rapidly disseminated by the GRB Coordinate Network (GCN), resulting
in multiwavelength followup observations and frequently burst
redshifts.  However, the BAT's 15--150~keV energy band is often
insufficient to determine the spectrum of the prompt burst emission,
particularly $E_p$, the `peak energy' where most of the burst energy
is radiated (see \S~\ref{sec:Phenomenological}); $E_p$ is important
not only for burst energetics but also for reported relationships
between intrinsic burst parameters
\citep{Amati:06,Ghirlanda:04,Firmani:06} that may turn bursts into
standard candles. And for those bursts where the BAT can determine
$E_p$, {\it Swift} cannot determine whether there is a second
emission component above the 15--150~keV band (as discussed in
\S\ref{sec:Theory}). In addition, {\it Swift}'s burst afterglow
observations `only' extend to the X-ray band ($E<10$~keV); as
discussed in \S~\ref{sec:Previous_Observations}, EGRET detected
GeV-band prompt and afterglow emissions \citep{Hurley:94,Dingus:03}.
Thus $Fermi$ and {\it Swift} capabilities complement each other
\citep{Stamatikos:08a}; between the UVOT, XRT, BAT, GBM and LAT, the
two mission's observations span 11 energy decades.

The $Fermi$ and {\it Swift} missions are working to increase the
number of bursts that are observed simultaneously by the BAT and the
LAT; this will increase the number of bursts with localizations,
redshifts, spectra and optical through gamma-ray afterglows.
Simultaneous burst observations by $Fermi$ and other burst missions
(e.g., {\it AGILE}, {\it INTEGRAL}, {\it Konus}-Wind, {\it RHESSI},
{\it Suzaku}-WAM) will also complement each other and permit
cross-calibration, but {\it Swift}'s pointing is the most flexible
\citep{Band:08}.

$Fermi$'s and {\it Swift}'s low earth orbits (altitudes of $\sim$565
and $\sim$590~km, respectively) are inclined to the Earth's equator
by 25.6$^\circ$ and 20.6$^\circ$, respectively.  The two orbits will
beat with a period of $\sim$13 days, that is, the two missions will
be on the same side, or opposite sides, of the Earth with a nearly
two week period.  Because of the uniformity of the LAT's
sky-exposure and the large FOVs of the BAT and the LAT, the relative
inclination of the two orbits (which can be as small as 5$^\circ$ or
as large as 46$^\circ$) has little effect on the overlap of the
FOVs. The relative inclination varies with a period of approximately
6.5 years.

In general $Fermi$ will survey the sky, pointing the LAT 35$^\circ$
above or below the orbital plane (as described in
\S~\ref{sec:Fermi_Mission}). On the other hand, every orbit {\it
Swift} points the Narrow-Field Instruments (NFIs---the XRT and UVOT)
at a number of targets that satisfy the mission's observational
constraints: the NFIs cannot be pointed near the Sun, moon, horizon
or ram direction; anti-Sun observations are preferred to increase
the detection of bursts during Earth's night. Since $Fermi$'s
observing mode will not change, but {\it Swift}'s timeline is by
design extremely flexible, increasing the overlap between the
mission's FOVs, and thus increasing the number of simultaneous burst
detections, will be done through {\it Swift}'s targeting.  Between
following-up bursts the {\it Swift} NFIs are used for other
observation programs (and will observe $Fermi$ sources).  By choosing
NFI targets at times that will increase the LAT-BAT overlap, we
estimate that this overlap can be improved by a factor of $\sim$2
without sacrificing {\it Swift}'s science objectives. Note that
increasing the BAT-LAT overlap will by necessity increase the
overlap between the BAT and GBM.

{\it Swift} detects $\sim$100 bursts per year, and {\nicola approximately one LAT detection per month is} 
anticipated, although this prediction of the LAT's detection rate is based on
extrapolations from lower energy (see
\S~\ref{sec:Sensitivity}). Given the differences in the
detectability of typical bursts, we assume that {\it
Swift}'s BAT will detect all the bursts that the LAT will
detect when the burst is in both their FOVs
The LAT's larger FOV compensates for the BAT's greater ability to
detect typical bursts, resulting in comparable detection
rates. Based on a number of modeling assumptions, and
assuming that {\it Swift}'s targeting can increase the
overlap of the BAT and LAT FOVs by $\times$2, we estimate
$\sim${\nicola 10} 
BAT bursts per year with LAT detections or upper
limits, and $\sim${\nicola 4} 
LAT bursts per year with BAT detections.  
We emphasize that our estimates of the LAT
detection rate assumes that the 10--1000~keV component
observed by BATSE, BAT and now the GBM extrapolates
unbroken into the LAT's energy band.

\subsection{TeV Observations}\label{sec:TeV}

The synergy between $Fermi$ and ground-based telescopes operating
above a few tens of GeV will expand the study of the still-unknown
spectral and temporal properties of GRBs above a few GeV. Extending
the analysis of burst temporal and spectral properties to even
higher energies would have a large impact on the knowledge of the
particle acceleration and emission processes occurring in the burst
environment.  High energy spectra would probe the distant Universe,
revealing the universe's transparency to high-energy gamma-rays and
measuring EBL. The requirements for a good coordination of $Fermi$
with TeV observatories are quite simple, and we examine the
potential of such simultaneous observations in terms of expected
rates of alerts and sensitivity.

Major TeV observatories operate above $\sim$100~GeV (or somewhat
lower for the next generation of instruments), and Imaging
Atmospheric Cherenkov Telescopes (IACTs) have a sensitivity of
$10^{-11}$ to $10^{-9}$~erg~cm$^{-2}$ to the latter part of the
prompt phase and early afterglow emission of GRBs (i.e., from
$\sim$10~s to a few hours after the trigger time). The
observatories' duty cycle, FOV and sky coverage will determine their
response to $Fermi$ alerts. With a high duty cycle ($\sim$100\%) and a
good sky coverage ($\sim$20\%), ground arrays like MILAGRO and ARGO
will be able to react to any alert provided by the GBM or the LAT.
In contrast, IACTs like CANGAROO, HESS, MAGIC, VERITAS, or STACEE
have a low duty cycle ($\sim$10\%) because they observe only during
clear and moonless nights, but they can slew to any location within
a few minutes and access $\sim$20\% of the sky. Because of their
small FOV ($\sim$5$^\circ$), IACTs will require a GRB position
accuracy of $\pm$1$^\circ$ and thus will respond effectively to LAT
alerts only.

Using a phenomenological model to describe GRB properties in the LAT
range, 
{\nicola we combine the estimated GRB detection rate (1 GRB per month) with the above duty cycle and sky coverage to compute the possible joint observations by $Fermi$ and TeV
experiments.} 
$Fermi$ should provide $\sim$40 alerts (including 2 to 5
LAT alerts) per year during the prompt burst phase, that ground
arrays will be able to follow up. Few of them will
be followed-up by IACTs due to localization accuracy and to
observing time constraints.{\nicola The LAT detected bursts per year suitable for TeV followup} should be considered as the highest priority targets in TeV telescope plans. 
A few afterglows per year may be also followed-up by IACTs, while ground
arrays will probably be much less sensitive to afterglows.

\subsection{Neutrino Observations}\label{sec:neutrino}

A major step forward in understanding of the
microphysics of the GRB central engines might be achieved via the
detection of non-electromagnetic emission such as
gravitational waves \citep{Abbott:05} and neutrinos. Because they are
weakly-interacting, neutrinos are unique (albeit
elusive) cosmic messengers because they are not absorbed nor deflected on their way to the observer. The viability of high energy neutrino
astronomy \citep{Gaisser:95} opens a new observing channel
that complements the high energy electromagnetic spectrum
that will be probed directly by the LAT.

Hadronic fireball models (\S\ref{sec:Leptonic_Hadronic}), predict a
taxonomy of correlated MeV to EeV neutrinos of varying flavor and
arrival times. Ideal for detection are $\sim$TeV-PeV muon neutrinos
\citep{Waxman:97a} produced as the leptonic decay products of
photomeson interactions $\left(p+\gamma \rightarrow \Delta ^{+}
\rightarrow \pi ^{+} + [n] \rightarrow \mu^{+}+\nu_{\mu} \rightarrow
e^{+} + \nu_e + \bar{\nu}_{\mu}+\nu_{\mu}\right)$ within the
internal shocks of the relativistic fireball. Since the prompt gamma
rays act as the ambient photon target field, the burst neutrinos are
expected to be spatially and temporally coincident with the
gamma-ray emission. Therefore Antarctic Cherenkov telescopes such as
Antarctic Muon and Neutrino Detector Array (AMANDA)
\citep{Ahrens:02} and IceCube \citep{Ahrens:04} can perform a nearly
background-free search for burst neutrinos correlated with the
prompt gamma-ray emission \citep{Stamatikos:05,Stamatikos:06}.
Neutrino telescopes have FOVs determined by their position on the
Earth, and accumulate and preserve their data, and therefore need
not to respond to bursts in realtime.  Instead, the neutrino data
archived is searched periodically for neutrinos correlated with the
time and position of prompt burst emission. Analysis of AMANDA data
has resulted in the most stringent upper limits upon correlated
multi-flavored neutrino emission from GRBs
\citep{Achterberg:07,Achterberg:08}. AMANDA's km-scale successor,
IceCube, is currently under construction with anticipated completion
by $\sim$2010, and thus will operate during the $Fermi$ era.

\section{Conclusions and Future
Work}\label{sec:Conclusions}

In this paper we provided an overview of the LAT's capabilities to
reveal the rich burst phenomenology in the $>$100~MeV band at which
the EGRET observations merely hinted, and which theoretical
scenarios predict.  These capabilities can be realized only through
efficient analysis techniques and software.  In this final section
we discuss the future analysis development that we anticipate during
the early part of the $Fermi$ mission.

Burst triggers are applied to the LAT data both onboard and
on-ground.  The onboard trigger contends with a higher
non-burst background rate, but can provide burst
notifications and localizations within tens of seconds
after the burst, while the on-ground trigger is more
sensitive because the background can be reduced, but the
burst notification and localizations have a $\sim$3 hr
latency.  The thresholds for both triggers depend on the
actual instrument response and background rates that are
only now being evaluated. Thus during the mission's early
phase we will tune the detection algorithms to minimize
false triggers and maximize the detection sensitivity.

In particular, we are investigating various `cuts' of the
reconstructed events used by the on-ground detection
algorithms. These cuts do not merely increase or decreased
the effective area and the background rate, but also change
their energy dependence. Relative changes in the effective
area and background rate affect the detectability of bursts
of different durations, since the background is less
important for detecting short bursts.

The GBM and LAT spectra will be analyzed jointly, giving
spectral fits from $\sim$8~keV to over 300~GeV, a bandpass
of up to 7.5 energy decades.  Typically the spectral
analysis will fit the parameters of functional forms such
as the `Band' function.

However, given the theoretical uncertainties in the
underlying GRB spectrum in the LAT band (e.g., the unknown
high energy attenuation by the EBL and intrinsic photon
fields), we will explore model-independent spectral
reconstruction. Deconvolution of instrument response
effects in the Poisson statistics regime is notoriously
difficult, but there have been advances in recent years.
For example, \citet{Nowak:00} derived a Bayesian multiscale
framework that is inspired by wavelet methods, but adapted
for Poisson statistics; using these methods, they
reconstructed a Solar flare emission line spectrum observed
by {\it CGRO}'s COMPTEL. \citet{DAgostini:95} derived
another Bayesian iterative method for deconvolving spectra;
uncertainties on the unfolded distribution can be estimated
from a covariance matrix.

Thus we anticipate an exciting mission exploring new burst
phenomena and developing the techniques to extract the
maximum information from the LAT.

\begin{acknowledgements}

We dedicate this paper to the memory of our colleague David Band, who died March 16 2009.  His contributions to the the field of GRB spectroscopy cannot be overestimated.   He played a large role in the fruition of GRB science goals promised in this paper, and realised following the launch of Fermi.  His presence on the Fermi team is already greatly missed.

We thank the members of the LAT instrument team, GBM instrument team
and the $Fermi$ Project for their exceptional efforts in developing
the $Fermi$ observatory.  M.~Stamatikos is supported by an NPP
Fellowship at NASA-GSFC administered by ORAU.

The $Fermi$ LAT Collaboration acknowledges support from a number of agencies and institutes for both the development and the operation of the LAT as well as scientific data analysis.  These include the National Aeronautics and Space Administration and the Department of Energy in the United States, the Commissariat \`a l'Energie Atomique and the Centre National de la Recherche Scientifique / Institut National de Physique Nucl\'eaire et de Physique des Particules in France, the Agenzia Spaziale Italiana and the Istituto Nazionale di Fisica Nucleare in Italy, the Ministry of Education, Culture, Sports, Science and Technology (MEXT), High Energy Accelerator Research Organization (KEK) and Japan Aerospace Exploration Agency (JAXA) in Japan, and the K.~A. Wallenberg Foundation, the Swedish Research Council and the Swedish National Space Board in Sweden.
Additional support from the Istituto Nazionale di Astrofisica in Italy for science analysis during the operations phase is also gratefully acknowledged.

\end{acknowledgements}

\bibliography{LAT_GRB}

\end{document}